\documentclass[preprint2]{emulateapj}

\newcommand{\chn}{{\it Chandra}}

\newcommand{\sample}{93}

\usepackage{graphicx}
\usepackage{longtable}
\usepackage{mdwlist}

\usepackage{natbib}
\usepackage{multirow}
\usepackage{upgreek}
\usepackage[colorlinks=true, pdfstartview=FitV, linkcolor=blue,citecolor=blue, urlcolor=blue]{hyperref}
\usepackage{wasysym}
\usepackage{txfonts}
\usepackage{gensymb}
\usepackage{morefloats}
\usepackage{threeparttable}
\usepackage{rotating}
\usepackage{graphicx}

\newcommand{\noprint}[1]{}

\slugcomment{version \today: fm} 

\shorttitle{The \chn\ 3CR extragalactic survey}
\shortauthors{F. Massaro et al.  2015}

\begin{document}
\title{The Chandra survey of extragalactic sources in the 3CR
catalog:\\ X-ray emission from nuclei, jets and hotspots in the
Chandra archival observations} 

\author{
F. Massaro\altaffilmark{1,2}, 
D. E. Harris\altaffilmark{3},
E. Liuzzo\altaffilmark{4},
M. Orienti\altaffilmark{4},
R. Paladino\altaffilmark{4,5},\\
A. Paggi\altaffilmark{3},
G. R. Tremblay\altaffilmark{2},
B. J. Wilkes\altaffilmark{3},
J. Kuraszkiewicz\altaffilmark{3},
S. A. Baum\altaffilmark{6,7}
\&
C. P. O'Dea\altaffilmark{6,8}
}

\altaffiltext{1}{Dipartimento di Fisica, Universit\`a degli Studi di Torino, via Pietro Giuria 1, I-10125 Torino, Italy.}
\altaffiltext{2}{Yale Center for Astronomy and Astrophysics, Physics Department, Yale University, PO Box 208120, New Haven, CT 06520-8120, USA.}
\altaffiltext{3}{Smithsonian Astrophysical Observatory, 60 Garden Street, Cambridge, MA 02138, USA.}
\altaffiltext{4}{Istituto di Radioastronomia, INAF, via Gobetti 101, 40129, Bologna, Italy.}
\altaffiltext{5}{Department of Physics and Astronomy, University of Bologna, V.le Berti Pichat 6/2, 40127 Bologna, Italy}
\altaffiltext{6}{University of Manitoba,  Dept of Physics and Astronomy, Winnipeg, MB R3T 2N2, Canada}
\altaffiltext{7}{Carlson Center for Imaging Science 76-3144, 84 Lomb Memorial Dr., Rochester, NY 14623, USA}
\altaffiltext{8}{School of  Physics and Astronomy, Rochester Institute of Technology,  84 Lomb Memorial Dr., Rochester, NY 14623, USA}

\begin{abstract} 
As part of our program to build a complete radio and X-ray database 
of all the 3CR extragalactic radio sources, we present an analysis of 93 sources 
for which \chn\ archival data are available. 
Most of these sources have been already published. 
Here we provide a uniform re-analysis and present nuclear X-ray fluxes and X-ray emission associated with
radio jet knots and hotspots using both publicly available radio images and new
radio images that have been constructed from data available in the VLA
archive. For about 1/3 of the sources in the selected sample
a comparison between the \chn\ and the radio observations was 
not reported in the literature: we find X-ray detections of 2 new radio jet knots and 17 hotspots. 
We also report the X-ray detection of extended emission 
from the intergalactic medium of 15 galaxy clusters, two of which
were most likely unknown previously.
\end{abstract}

\keywords{galaxies: active --- X-rays: general --- radio continuum: galaxies}

\section{Introduction}
\label{sec:intro}

The first release of the Third Cambridge catalog (3C), performed at 159 MHz, was published in 1959 \citep{edge59}.
In 1962 Bennett et al. revised the whole 3C catalog using observations at 178 MHz and this revised version (3CR) was considered 
a definitive list of the brightest radio sources in the Northern Hemisphere for many years. 
The flux limit of the 3CR catalog is set to 9 Jy at 178 MHz and it covers the whole Northern Hemisphere above -5\degr\ in Declination.
Then, in 1985, Spinrad, Djorgovski, Marr and Aguilar presented the last revised version of the Third Cambridge catalog (3CR) \citep{bennett62}
listing 298 extragalactic radio sources \citep[see also][]{edge59,mackay71,smith76,smith80}
including new revised positions, redshifts and magnitudes  
having 91\% of the sources out of the Galactic plane (i.e., Galactic latitude $|b|>$10\degr).
Since then several photometric and spectroscopic surveys have been carried out 
to obtain multifrequency coverage of the 3CR catalog.  
All the 3CR sources at redshift $z<$0.3 have been already observed with the
Hubble Space Telescope (HST) \citep[e.g.,][]{chiaberge00,tremblay09} while a near infrared, optical and ultraviolet survey 
for higher redshift sources is still ongoing.
A large fraction of the 3CR radio sources were also targets of the spectroscopic survey carried 
out with the Telescopio Nazionale Galileo \citep[TNG; e.g.,][]{buttiglione09}. 
Radio images with arcsecond resolution for the majority of the 3CR sources are available
from the NRAO Very Large Array (VLA) Archive Survey (NVAS)\footnote{\underline{http://archive.nrao.edu/nvas/}}
and from the MERLIN archive\footnote{\underline{http://www.jb.man.ac.uk/cgi-bin/merlin\_retrieve.pl}}.
As a radio low frequency catalog, the selection criteria for the 3CR are unbiased with
respect to X--rays. Since it spans a wide range of redshift and radio power
and has a vast multifrequency database of ground and spaced based observations 
for comparison, it is an ideal sample to investigate properties of active galaxies.

Motivated by the large number of multifrequency observations already
available for the 3CR sources, we have undertaken a project to ensure that each 3CR extragalactic source
has at least an exploratory/snapshot \chn\ observation.  
We have chosen to achieve this goal in a step wise strategy, working out in redshift with modest
proposals each cycle to minimize the impact on the \chn\ schedule.
A description of our progress in this endeavor is given in the following sections.
 
 In this paper we present the X-ray analyses of most of the 3CR
sources present in the \chn\ archive which have not already been published
by us with our standard procedures:  i.e. the snapshot surveys
\citep{massaro10,massaro12,massaro13,massaro15} and the 3CR sources in the XJET
sample \citep{massaro11}. Our main aim is to provide a uniform analysis for
all the archival observations. X-ray flux maps were constructed and compared with
radio images to search for any X-ray emission associated with radio
jet knots, hotspots and lobes. In some cases new radio images have been been
constructed from archival VLA data for comparison with the X-ray images.  
We report the measurements of the
X-ray nuclear emission for all sources in our sample, but we did not perform a detailed spectral analysis because
most of them (i.e., $>$70\%) have already been reported in the literature \citep[see e.g.,][]{hardcastle09,balmaverde12,wilkes13,kuraszkiewicz15}.

The paper is organized as follows. A brief historical overview of the \chn\ observations of the 3CR sources
is provided in \S~\ref{sec:history} while the description of the selected
sample is presented in \S~\ref{sec:sample}.  Data reduction procedures
are given in \S~\ref{sec:obs} while results are discussed in
\S~\ref{sec:results}.  Then, \S~\ref{sec:summary} is devoted to our
summary and conclusions. 
Finally, in the Appendix, we show the X-ray images with radio contours
superposed for all the sources
analyzed (\S~\ref{sec:images}) and a summary of the \chn\ observations 
for the entire sample of 3CR extragalactic sources (\S~\ref{sec:state}).

For numerical results, cgs units are used
unless stated otherwise and a flat cosmology was assumed with $H_0=72$
km s$^{-1}$ Mpc$^{-1}$, $\Omega_{M}=0.27$ and $\Omega_{\Lambda}=0.73$
\citep{dunkley09}, to be consistent with our previous analyses
\citep[e.g.,][]{massaro10,massaro12,massaro13}.  Spectral indices,
$\alpha$, are defined by flux density, S$_{\nu}\propto\nu^{-\alpha}$.

\section{History of the 3CR Chandra Survey}
\label{sec:history}

A large fraction of the X--ray studies of 3CR extragalactic sources 
observed with \chn\ is biased towards observations of ``favorite'' X--ray
bright sources or objects with well-known interesting features and/or
peculiarities (e.g. sources in the center of bright galaxy clusters)
rather than consisting of well defined samples.
However to complete the X-ray coverage for the whole 3CR catalog and
to obtain a complete and uniform multifrequency database of these extragalactic radio sources, 
we started, during \chn\ Cycle 9, an X-ray snapshot survey 
of the 3CR sources previously unobserved by \chn.
Several subsets of the 3CR sample have been observed by other groups \citep[e.g.,][]{wilkes13,kuraszkiewicz15}.


The 3CR extragalactic catalog includes 298 sources with 248 of them already in the \chn\ archive.
Among the observed ones we have already published 47 sources as part of the XJET 
project \citep{massaro11}\footnote{\underline{http://hea-www.cfa.harvard.edu/XJET/}}
and an additional 98 as part of our 3CR \chn\ snapshot survey \citep{massaro10,massaro12,massaro13,massaro15}.
Here we publish an additional 93 sources from the \chn\ archive.
It is worth noting that of the remaining 50 sources unobserved by \chn, half are unidentified, i.e. lacking of an assigned optical
counterpart and thus unclassified.
Table~\ref{tab:summary} gives the references for the 145 sources we have already processed and published.
\begin{table*}
\tiny
\begin{center} 
\caption{Summary of the 3CR sources analyzed in our previous investigations}
\label{tab:summary}
\begin{tabular}{|lcccll|}
\hline
Program  & Cycle & Proposal Number & Number of sources   & Redshift range & Reference \\
\hline 
\noalign{\smallskip}
3CR snapshot survey & 9 & 09700745 & 30$^*$ & $z<$0.3 & Massaro et al. (2010) \\
XJET$^+$ & --- & --- & 47 & --- & Massaro et al. (2011) \\
3CR snapshot survey & 12 & 12700211 & 26 & $z<$0.3 & Massaro et al. (2012) \\
3CR snapshot survey & 13 & 13700190 & 19 & $z<$0.5 & Massaro et al. (2013) \\
3CR snapshot survey & 15 & 15700111 & 23 & $z<$1.0 & Massaro et al. (2015)\\
Archival project$^+$ & --- & --- & 93 & --- & This work \\
\noalign{\smallskip}
\hline
\end{tabular}\\
\end{center}
$^*$ The AO9 sample includes 3CR 346 that was re-observed in Cycle
12 because during Cycle 9 its \chn\ observation was affected by high
background \citep[see][for details]{massaro10}.\\
$^+$ The redshift ranges for both the archival and the XJET samples are
unbounded w.r.t. selection.
\end{table*} 

According to the redshift estimates reported in the 3CR catalog,
the \chn\ archive now contains all the 3CR sources up to $z=$0.5 (i.e., 150 sources),
with the only exceptions of: 
3CR 27, at $z=$0.184, 3CR 69 at $z=$0.458 \citep{hiltner91} and 3CR 93 at $z=$0.357, 
as confirmed by Ho \& MinJin (2009).

In this paper of the series we present the X-ray analysis of the 3CR sources that were not
described either in our XJET database \citep{massaro11} 
or in the analyses of the \chn\ snapshot survey (Massaro et al. 2010, 2012, 2013).
This will permit us to obtain a uniform database of the 3CR X-ray and radio observations.

\section{Sample selection for 3CR archival observations}
\label{sec:sample}
In the present paper we uniformly analyzed 93 3CR sources observed 
by \chn\ that were not reported in our previous investigations. 
We excluded from the present archival analysis seven 
3CR sources which have been
extensively discussed in the literature and which have an accumulated
exposure time greater than 80 ks each.  The excluded sources are: 
3CR 66A \citep[e.g., ][]{abdo11}, 
3CR 71 \citep[alias NGC 1068; e.g., ][]{brinkman02}, 
3CR 84 \citep[alias NGC1275 or Perseus A; e.g., ][]{fabian03}, 
3CR 186 \citep{siemiginowska10}, 
3CR 231 \citep[alias M82; e.g., ][]{griffiths00}, 
3CR 317 \citep[alias Abell 2052; e.g., ][]{blanton09} 
and 3CR 348 \citep[alias Hercules A; e.g., ][]{nulsen05}. 
In addition we also did not select for our analysis the following three cases:
3CR 236, 3CR 326, 3CR 386 since the PI of these observations 
is currently working on them (M. Birkinshaw, priv. comm.).

In Table~\ref{tab:log1} and \ref{tab:log2}, we list all the selected \sample\ sources, their coordinates,
redshift estimates, luminosity distance, the \chn\ observation ID number, exposure times
and observing dates. In the same tables we also list the references where the \chn\ observations
were analyzed/presented. 
%
%
%
%

\section{Data reduction and data analysis}
\label{sec:obs}

The radio and X-ray data reduction and analysis procedures adopted 
in the present analysis were extensively described 
in Massaro et al. (2012, 2013) and references therein. Here we report only the basic details.

\subsection{Radio observations}
\label{sec:radio}

Radio observations presented in this paper were retrieved from
publicly available websites of M. J. Hardcastle and C. C. Cheung,
from the NVAS (National Radio Astronomy Observatory VLA Archive
Survey), from NED (NASA Extragalactic Database), from the DRAGN
website, or were constructed from data available in the VLA archives.
Summary of the archival data used is reported in Table~\ref{tab:radio}.
In the latter case, to produce our final images, we calibrated the
data with standard procedures using AIPS (Astronomical Image
Processing System),  edited the visibilities, and carried out
a few self-calibration cycles.
Image parameters for each figure are given in the Appendix.

\subsection{X-ray observations}
\label{sec:xray}

The data reduction was performed following the standard procedure
described in the \chn\ Interactive Analysis of Observations (CIAO)
threads\footnote{http://cxc.harvard.edu/ciao/guides/index.html}, using
CIAO v4.6 and the \chn\ Calibration Database (CALDB) version 4.6.2.
Level 2 event files were generated using the $acis\_process\_events$
task and events were filtered for grades 0,2,3,4,6.  
Lightcurves were also extracted for every dataset thus confirming the 
absence of to verify the absence of high background intervals.
Astrometric registration was achieved by
aligning the nuclear X-ray position with that of the radio \citep[see
e.g.][]{massaro10,massaro11}.

Three different flux maps were created in the energy ranges: 0.5 -- 1
keV (soft), 1 -- 2 keV (medium), 2 -- 7 keV (hard).
Flux maps, as implemented in CIAO, are corrected for exposure time 
and effective area and our implementation
used monochromatic exposure maps.  
Each band is assigned a nominal energy; in
our case the nominal energies are 0.75,
1.4, and 4 keV for the soft, medium and hard band, respectively 
and the exposure maps are constructed for these nominal values. 
Since the natural units of X-ray flux maps are counts/sec/cm$^2$
we converted them to cgs units by multiplying each event by 
the nominal energy of its band, thereby assuming that every event in the band has the same energy.  
However, when we perform our photometry, we make the necessary correction to recover the
observed erg/cm$^2$/s. The use of the ``nominal energy" is only to get the correct units.
The total energy for {\bf any} particular region is recovered by
applying a correction factor of $E(average)/E(nominal)$ to the photometric measurement
to derive $E(average)$, the actual values were measured.
This correction ranged from a few to 15\%.  

To measure observed fluxes for the nuclear emission as well as for any
feature, a region of size and shape appropriate to the observed X-ray emission was chosen.
Two background regions, each with the same shape and size, were chosen
so as to avoid emission from other parts of the source
and to sample both sides of jet features or two areas close to
hotspots.  The flux in any particular band for any particular region
was measured using funtools\footnote{http://www.cfa.harvard.edu/$\sim$john/funtools}
\citep[see also][]{massaro11}.  

A one $\sigma$ error is calculated based
on the usual $\sqrt{number-of-counts}$ in the source and background
regions. Fluxes reported here are not corrected for the Galactic absorption.
X-ray fluxes measured for the cores are reported in
Table~\ref{tab:cores1} and \ref{tab:cores2} while those for the radio jet knots and hotspots
detected are given in Table~\ref{tab:features}.

At the focal point, the Chandra mirrors produce an image of a point source with a FWHM 
of the order 0\arcsec.7.   Since the native ACIS pixel size is 0\arcsec.492, the data are undersampled.  
To recover the resolution inherent in the telescope, we normally regrid our images
with a binning factor of 1/2, 1/4, or 1/8 of the native ACIS pixel size.  
The choice of binning factor was dictated by the angular size of the
radio source and by the number of counts in source components.
The fact that the telescope dithers during each observation, 
together with the fact that real numbers rather than integers 
are used throughout for event location, permits us to achieve adequate Nyquist sampling of the point spread function (PSF).
For sources of large angular extent 1/2 or
no regridding was used \citep[see also][for more
details]{massaro12,massaro13}.

\section{Results}
\label{sec:results}
X-ray emission was clearly detected for 85 out of 93 nuclei in our sample.
For 3CR 441 we did not perform X-ray photometry since 
the number of counts measured within a circular region of 2$\arcsec$ centered on the radio position
is consistent with the background.
For an additional four sources, namely: 3CR 99, 3CR 220.3, 3CR 256 and 3CR 368 we measured too few X-ray counts to
define a discrete nucleus in the \chn\ image.
In the three sources 3CR 28, 3CR 288, and 3CR 310 
we could not measure the X-ray flux because the extended emission from the cluster washes out the discrete nuclear emission.
For all the other sources the nuclear X-ray fluxes in the three bands (see \S~\ref{sec:xray}) together with
their X-ray luminosities are reported in Tables~\ref{tab:cores1} and \ref{tab:cores2}.  

A detailed spectral analysis for the bright cores is beyond the scope of
this paper since a large fraction of the sources were
extensively analyzed in the literature.  
As done in our previous investigations, in Tables~\ref{tab:cores1} and \ref{tab:cores2} we also report
an 'extended emission' parameter computed as 
the ratio of the net counts in the r\,=\,2\arcsec\ circle to the net counts in the
r\,=\,10\arcsec\ circular region surrounding the core of each 3CR source (i.e., Ext. Ratio ``Extent Ratio'').  
Values significantly less than 0.9 indicate the presence of extended emission around the nuclear
component \citep[e.g.,][]{massaro10,massaro13}.

We detected and report here, the X-ray emission of 8 radio jet
knots in 7 sources and 17 hotspots in 13 objects while no emission arising from
lobes was found. To the best of our knowledge, 2 of our jet
knot detections (3CR 78 and 3CR 245) and all the hotspots have not previously
been reported in the literature.

X-ray fluxes for radio jet knots and hotspots found in the 3CR sample are reported in Table~\ref{tab:features}, 
where the classification of each component is also provided.
The significance of all detections is above 5$\sigma$, with the exception of
the northern hotspot in 3CR 470 (i.e., n14.4) that corresponds to a $\sim$1$\sigma$ detection.
These significances have been computed assuming a Poisson distribution for the background 
as done in Massaro et al. (2013).

In our sample there are also 15 sources, members of galaxy clusters,
for which extended X-ray emission is clearly visible, 13 of these were previously known as cluster related X-ray sources
while the remaining 2, namely 3CR\,427.1 and 3CR\,449, are reported here for the first time
to the best of our knowledge.
For each galaxy cluster in our sample we present the basic parameters in Table~\ref{tab:clusters}:
the associated 3CR radio source, the alternative X-ray or optical name if it was a known galaxy cluster,
the size of the X-ray emission estimated as the radius of a circular region surrounding its emission, 
both in arcseconds and in kpc, together with the number of counts within the same area.
A dedicated analysis of the 3CR sources in galaxy clusters, listed in the \chn\ snapshot survey, will be presented in a future paper.

All the X-ray images for the selected sample are presented in the Appendix.

\section{Summary and Conclusions}
\label{sec:summary}

We have described the combined radio-X-ray analyses of 93 3CR
radio sources for which \chn\ observations requested by others for
many different reasons, were already present in the archive.  The main
objectives of the present analysis are: (1) to present a uniform X-ray
and radio database for the 3CR catalog, (2) to search for possible
detections of X-ray emission from radio jet knots, hotspots and lobes and (3)
to look for new galaxy cluster detections surrounding the 3CR radio
sources.

In order to perform the radio--X-ray comparison we reduced
archival radio observations for 6 sources. 
We focused on the comparison between the radio and the X-ray emission
from extended components such as radio jet knots, hotspots, and lobes.
We discovered 2 new radio jet knots and 17 hotspots emitting in the X-rays.  
Flux maps for all the X-ray observations were
constructed and we provided photometric results for all the extended
components detected.

All the radio knots and hotspots have been classified on the basis of the
radio morphology of their parent source, adopting the definition
suggested by Leahy et al. (1997) for the hotspots, i.e., brightness
peaks which are neither the core nor a part of the jet, usually lying
where the jet terminates, and considering all other discrete
brightness enhancements as jet knots.

The following conventions for labeling the extended structures
detected in the X-rays was adopted.  We indicated with the letter
`k' the jet knots and with `h' the hotspots; then the name of each
component is a combination of one letter (indicating the cardinal
direction of the radio feature with respect to the nucleus) and one
number (indicating the distance from the core in arcsec)
as described in Massaro et al. (2011).
We also reported the presence of 15 X-ray galaxy clusters associated with
the selected 3CR source, 13 already known in the X-rays and 2, namely 3CR\,427.1 and 3CR\,449, 
reported here for the first time to the best of our knowledge.  

In the Appendix we present X-ray images with radio contours for all the 93 sources 
analyzed in this paper (\S~\ref{sec:images}) and in (\S~\ref{sec:state}) we give 
the \chn\ status of the observations for all extragalactic 3CR sources.

\acknowledgments 
We thank the anonymous referee for useful comments that led to improvements in the paper.
We are grateful to M. Hardcastle and C. C. Cheung for providing
several radio images of the 3CR sources while the remaining ones were
downloaded from the NVAS\footnote{http://archive.nrao.edu/nvas/} (NRAO
VLA Archive Survey), NED\footnote{http://ned.ipac.caltech.edu/} (Nasa
Extragalactic Database) and from the DRAGN
webpage\footnote{http://www.jb.man.ac.uk/atlas/}.
This investigation is supported by the NASA grants GO1-12125A,
GO2-13115X, and GO4-15097X.  G.R.T acknowledges support by the
European Community's Seventh Framework Programme (/FP7/2007- 2013/)
under grant agreement No. 229517.  This work was also supported by
contributions of European Union, Valle DÕAosta Region and the
Italian Minister for Work and Welfare. 
The National Radio Astronomy Observatory is operated by Associated Universities, Inc.,
under contract with the National Science Foundation.
This research has made use of data obtained from the High-Energy Astrophysics Science Archive
Research Center (HEASARC) provided by NASA's Goddard Space Flight Center; 
the SIMBAD database operated at CDS,
Strasbourg, France; the NASA/IPAC Extragalactic Database
(NED) operated by the Jet Propulsion Laboratory, California
Institute of Technology, under contract with the National Aeronautics and Space Administration.
TOPCAT\footnote{\underline{http://www.star.bris.ac.uk/$\sim$mbt/topcat/}} 
\citep{taylor05} for the preparation and manipulation of the tabular data and the images.
SAOImage DS9 were used extensively in this work
for the preparation and manipulation of the
images.  SAOImage DS9 was developed by the Smithsonian Astrophysical
Observatory. 

{Facilities:} \facility{VLA}, \facility{MERLIN}, \facility{CXO (ACIS)}

\clearpage
\begin{table*} 
\caption{Source List of the archival \chn\ 3CR radio sources}
\label{tab:log1}
\tiny
\begin{tabular}{|lllllrrrlrl|}
\hline
3CR  & R.A. (J2000) & Dec. (J2000) & z & kpc scale & D$_L$ & \chn\   & Obs. Date  & Data Mode & Exposure & References \\
name & (hh mm ss)   & (dd mm ss)   &   & (kpc/arcsec) & (Mpc) & Obs. and proposal IDs & yyyy-mm-dd & & (ksec) & \\ 
\hline 
\noalign{\smallskip}
    2.0 & 00:06:22.6   & -00:04:24.6  & 1.0374 & 7.999 &  6849.63 &  5617 (06700116) & 2005-07-28 & ACIS-S FAINT  & 16.93 & Miller et al. (2011) \\ 
   13.0 & 00:34:14.500 & +39:24:17.00 & 1.351  & 8.357 &  9528.43 &  9241 (09700482) & 2008-06-01 & ACIS-S FAINT  & 19.53 & Wilkes et al. (2013) \\ 
   14.0 & 00:36:06.447 & +18:37:59.08 & 1.469  & 8.412 & 10578.07 &  9242 (09700482) & 2008-05-29 & ACIS-S FAINT  &  3.00 & Wilkes et al. (2013) \\ 
   22.0 & 00:50:56.222 & +51:12:03.26 & 0.936  & 7.792 &  6024.20 & 14994 (14700660) & 2013-06-05 & ACIS-S FAINT  &  9.35 & Kuraszkiewicz et al. (2015) \\ 
   28.0 & 00:55:50.6   & +26:24:36.7  & 0.1953 & 3.162 &   931.87 &  3233 (03800625) & 2002-10-07 & ACIS-I VFAINT & 49.72 & McCarthy et al. (2004), Donato et al. (2004) \\ 
   35.0 & 01:12:02.288 & +49:28:35.62 & 0.067  & 1.250 &   293.48 & 10240 (10700504) & 2009-03-08 & ACIS-I VFAINT & 25.63 & Isobe et al. (2011) \\
   40.0 & 01:26:00.616 & -01:20:42.44 & 0.018  & 0.356 &    75.99 &  7823 (08700576) & 2007-09-07 & ACIS-S VFAINT & 64.82 & Sun et al. (2009) \\
   43.0 & 01:29:59.776 & +23:38:19.85 & 1.459  & 8.409 & 10488.44 &  9324 (09700482) & 2008-06-17 & ACIS-S FAINT  &  3.04 & Wilkes et al. (2013) \\ 
   48.0 & 01:37:41.301 & +33:09:35.27 & 0.367  & 4.991 &  1923.58 &  3097 (03700781) & 2002-03-06 & ACIS-S VFAINT &  9.22 & Worrall et al. (2004) \\ 
   49.0 & 01:41:09.159 & +13:53:28.33 & 0.621  & 6.687 &  3837.59 & 14995 (14700660) & 2013-08-31 & ACIS-S FAINT  &  9.45 & Kuraszkiewicz et al. (2015) \\ 
   65.0 & 02:23:43.1   & +40:00:51.9  & 1.176  & 8.203 &  8011.89 &  9243 (09700482) & 2008-06-30 & ACIS-S FAINT  & 20.91 & Wilkes et al. (2013) \\ 
   68.1 & 02:32:28.8   & +34:23:45.9  & 1.238  & 8.269 &  8543.36 &  9244 (09700482) & 2008-02-10 & ACIS-S FAINT  &  3.05 & Wilkes et al. (2013) \\ 
   68.2 & 02:34:23.8   & +31:34:17.0  & 1.575  & 8.435 & 11537.52 &  9245 (09700482) & 2008-03-06 & ACIS-S FAINT  & 19.88 & Wilkes et al. (2013) \\ 
   75.0 & 02:57:41.570 & +06:01:36.92 & 0.0232 & 0.456 &    98.33 &  4181 (04800347) & 2003-09-19 & ACIS-I VFAINT & 21.49 & Balmaverde et al. (2006), Hudson et al. (2006) \\ 
   78.0 & 03:08:26.222 & +04:06:39.26 & 0.0287 & 0.560 &   122.12 &  4157 (04700407) & 2004-06-28 & ACIS-S VFAINT & 50.86 & Harwood \& Hardcastle (2012) \\
   88.0 & 03:27:54.171 & +02:33:42.24 & 0.0302 & 0.588 &   128.66 & 11977 (11800517) & 2009-10-06 & ACIS-S VFAINT & 49.62 & Sun et al. (2009) \\ 
   98.0 & 03:58:54.431 & +10:26:02.72 & 0.0305 & 0.594 &   129.97 & 10234 (10700504) & 2008-12-24 & ACIS-I VFAINT & 31.71 & Hodges-Kluck et al. (2010) \\ 
   99.0 & 04:01:07.6   & +00:36:33.1  & 0.426  & 5.474 &  2296.01 &  5680 (06700612) & 2005-11-28 & ACIS-S FAINT  &  5.07 & \\ 
  129.1 & 04:50:06.645 & +45:03:05.91 & 0.0222 & 0.436 &    94.03 &  2219 (02800530) & 2001-01-09 & ACIS-I FAINT  &  9.63 & Krawczynski et al. (2002) \\   
  136.1 & 05:16:03.275 & +24:58:25.68 & 0.064  & 1.198 &   279.75 &  9326 (09700606) & 2008-01-11 & ACIS-S FAINT  &  9.91 & Balmaverde et al. (2012) \\
  138.0 & 05:21:09.906 & +16:38:22.16 & 0.759  & 7.273 &  4641.94 & 14996 (14700660) & 2013-03-22 & ACIS-S FAINT  &  2.00 & Kuraszkiewicz et al. (2015) \\ 
  147.0 & 05:42:36.127 & +49:51:07.19 & 0.545  & 6.276 &  3271.83 & 14997 (14700660) & 2013-08-26 & ACIS-S FAINT  &  2.00 & Kuraszkiewicz et al. (2015) \\ 
  172.0 & 07:02:08.305 & +25:13:53.52 & 0.519  & 6.119 &  3083.34 & 14998 (14700660) & 2013-09-05 & ACIS-S FAINT  &  9.95 & Kuraszkiewicz et al. (2015) \\ 
  175.0 & 07:13:02.422 & +11:46:16.25 & 0.77   & 7.312 &  4725.35 & 14999 (14700660) & 2013-02-21 & ACIS-S FAINT  &  2.00 & Kuraszkiewicz et al. (2015) \\ 
  175.1 & 07:14:04.695 & +14:36:22.57 & 0.92   & 7.754 &  5896.15 & 15000 (14700660) & 2013-02-10 & ACIS-S FAINT  &  9.94 & Kuraszkiewicz et al. (2015) \\ 
  181.0 & 07:28:10.216 & +14:37:36.60 & 1.382  & 8.375 &  9802.28 &  9246 (09700482) & 2009-02-12 & ACIS-S FAINT  &  3.02 & Wilkes et al. (2013) \\
  184.0 & 07:39:24.4   & +70:23:10.0  & 0.994  & 7.917 &  6493.56 &  3226 (03800590) & 2002-09-22 & ACIS-S VFAINT & 18.89 & Belsole et al. (2004), Hardcastle et al. (2004) \\
  190.0 & 08:01:33.552 & +14:14:42.83 & 1.1956 & 8.225 &  8179.15 &  9247 (09700482) & 2007-12-31 & ACIS-S FAINT  &  3.06 & Wilkes et al. (2013) \\
  191.0 & 08:04:47.968 & +10:15:23.72 & 1.956  & 8.375 & 15096.31 &  5626 (06700234) & 2005-12-14 & ACIS-S VFAINT & 19.77 & Erlund et al. (2006) \\
  192.0 & 08:05:35.005 & +24:09:50.36 & 0.0597 & 1.123 &   260.18 &  9270 (09700606) & 2007-12-18 & ACIS-S FAINT  & 10.02 & Hodges-Kluck et al. (2010) \\
  196.0 & 08:13:36.058 & +48:13:02.66 & 0.871  & 7.627 &  5507.36 & 15001 (14700660) & 2013-03-23 & ACIS-S FAINT  &  2.00 & Kuraszkiewicz et al. (2015) \\ 
  200.0 & 08:27:25.384 & +29:18:45.01 & 0.458  & 5.711 &  2504.16 &   838 (01700549) & 2000-10-06 & ACIS-S FAINT  & 14.66 & Hardcastle et al. (2004) \\ 
  204.0 & 08:37:45.003 & +65:13:35.34 & 1.112  & 8.119 &  7470.55 &  9248 (09700482) & 2008-01-13 & ACIS-S FAINT  &  3.05 & Wilkes et al. (2013) \\
  205.0 & 08:39:06.534 & +57:54:17.09 & 1.534  & 8.429 & 11164.65 &  9249 (09700482) & 2008-01-26 & ACIS-S FAINT  &  9.67 & Wilkes et al. (2013) \\
  208.0 & 08:53:08.608 & +13:52:54.85 & 1.1115 & 8.118 &  7466.35 &  9250 (09700482) & 2008-01-08 & ACIS-S FAINT  &  3.01 & Wilkes et al. (2013) \\
  210.0 & 08:58:10.0   & +27:50:54.9  & 1.169  & 8.194 &  7952.27 &  5821 (06800802) & 2004-12-25 & ACIS-S VFAINT & 20.57 & Gilmour et al. (2009) \\
  215.0 & 09:06:31.874 & +16:46:11.81 & 0.4121 & 5.366 &  2206.92 &  3054 (03700563) & 2003-01-02 & ACIS-S FAINT  & 33.80 & Hartdcastle et al. (2004) \\
  216.0 & 09:09:33.498 & +42:53:46.51 & 0.6699 & 6.916 &  3978.18 & 15002 (14700660) & 2013-02-25 & ACIS-S FAINT  &  2.00 & Kuraszkiewicz et al. (2015) \\ 
  220.1 & 09:32:40.025 & +79:06:30.14 & 0.61   & 6.632 &  3545.74 &   839 (01700549) & 1999-12-29 & ACIS-S FAINT  & 18.92 & Worrall et al. (2001) \\
  220.3 & 09:39:23.4   & +83:15:26.2  & 0.68   & 6.961 &  4052.36 & 14992 (14700660) & 2013-01-21 & ACIS-S FAINT  &  9.94 & Haas et al. (2014)\\ 
  226.0 & 09:44:16.522 & +09:46:17.07 & 0.8177 & 7.471 &  5091.20 & 15003 (14700660) & 2013-10-07 & ACIS-S FAINT  &  9.94 & Kuraszkiewicz et al. (2015) \\ 
  228.0 & 09:50:10.794 & +14:20:00.68 & 0.5524 & 6.319 &  3141.17 &  2095 (02700363) & 2001-06-03 & ACIS-S FAINT  & 13.78 & Belsole et al. (2006) \\ 
  241.0 & 10:21:54.6   & +21:59:31.2  & 1.617  & 8.438 & 11921.71 &  9251 (09700482) & 2008-03-13 & ACIS-S FAINT  & 18.93 & Wilkes et al. (2013) \\ 
  245.0 & 10:42:44.609 & +12:03:31.15 & 1.0279 & 7.982 &  6771.29 &  2136 (02700500) & 2001-02-12 & ACIS-S FAINT  & 10.40 & Gambill et al. (2003) \\
  249.1 & 11:04:13.842 & +76:58:58.17 & 0.3115 & 4.474 &  1587.34 &  3986 (04700368) & 2003-07-02 & ACIS-I VFAINT & 24.04 & Stockton et al. (2006) \\
  252.0 & 11:11:32.995 & +35:40:41.50 & 1.1    & 8.102 &  7370.04 &  9252 (09700482) & 2008-03-11 & ACIS-S FAINT  & 19.45 & Wilkes et al. (2013) \\ 
\noalign{\smallskip}
\hline
\end{tabular}\\
Col. (1): The 3CR name.
Col. (2): Right ascension and Declination (equinox J2000) of the radio position used to perform the registration (see \S~\ref{sec:obs} for details). We reported here the original 3CR position \citep{spinrad85} of the sources for which the radio core was not clearly detected.
Col. (3): Redshift $z$. We also verified in the literature (e.g., NED and/or SIMBAD databases) if new $z$ values were reported after the release of the 3CR catalog.
Col. (4): The angular to linear scale factor in arcseconds. Cosmological parameters used to compute it are reported in \S~\ref{sec:intro}.
Col. (5): Luminosity Distance in Mpc. Cosmological parameters used to compute it are reported in \S~\ref{sec:intro}. 
Col. (6): \chn\ observation identification number. Proposal identification number is also reported in parenthesis.
Col. (7): \chn\ observation date.
Col. (8): Data mode indicates how the \chn\ ACIS detector was configured for the observation analyzed.
Col. (9): The total exposure.
Col. (10): The reference for the \chn\ observation.
\end{table*} 

\begin{table*} 
\caption{Source List of the archival \chn\ 3CR radio sources}
\label{tab:log2}
\tiny
\begin{tabular}{|llllrrrlrll|}
\hline
3CR  & R.A. (J2000) & Dec. (J2000) & z & kpc scale & D$_L$ & \chn\   & Obs. Date  & Data Mode & Exposure & References \\
name & (hh mm ss)   & (dd mm ss)   &   & (kpc/arcsec) & (Mpc) & Obs. and proposal IDs & yyyy-mm-dd & & (ksec) & \\ 
\hline 
\noalign{\smallskip}
  256.0 & 11:20:43.02  & +23:27:55.2  & 1.819  & 8.417 & 13798.51 &  1660 (02800089) & 2001-04-23 & ACIS-I VFAINT & 71.25 & Vikhlinin et al. (2002) \\ 
  263.1 & 11:43:25.094 & +22:06:56.10 & 0.824  & 7.490 &  5140.19 & 15004 (14700660) & 2013-03-20 & ACIS-S FAINT  &  9.94 & Kuraszkiewicz et al. (2015) \\ 
  266.0 & 11:45:43.30  & +49:46:08.0  & 1.275  & 8.302 &  8863.60 &  9253 (09700482) & 2008-02-16 & ACIS-S FAINT  & 18.23 & Wilkes et al. (2013) \\ 
  267.0 & 11:49:56.506 & +12:47:18.83 & 1.14   & 8.158 &  7706.55 &  9254 (09700482) & 2008-07-07 & ACIS-S FAINT  & 19.18 & Wilkes et al. (2013) \\ 
  268.1 & 12:00:24.482 & +73:00:45.81 & 0.97   & 7.868 &  6298.40 & 15005 (14700660) & 2013-07-08 & ACIS-S FAINT  &  9.94 & Kuraszkiewicz et al. (2015) \\ 
  268.3 & 12:06:24.89  & +64:13:37.9  & 0.3717 & 5.032 &  1952.68 & 10382 (10700678) & 2009-07-29 & ACIS-S VFAINT & 42.53 & \\
  268.4 & 12:09:13.610 & +43:39:20.89 & 1.4022 & 8.385 &  9981.44 &  9325 (09700482) & 2009-02-23 & ACIS-S FAINT  &  3.02 & Wilkes et al. (2013) \\ 
  270.1 & 12:20:33.881 & +33:43:11.99 & 1.5284 & 8.428 & 11113.88 &  9255 (09700482) & 2008-02-16 & ACIS-S FAINT  &  9.67 & Wilkes et al. (2012) \\ 
  277.1 & 12:52:26.353 & +56:34:19.58 & 0.3198 & 4.556 &  1636.83 &  3102 (03700781) & 2002-10-27 & ACIS-S VFAINT & 14.01 & Siemiginowska et al. (2008) \\ 
  277.3 & 12:54:12.010 & +27:37:33.86 & 0.0853 & 1.559 &   378.60 & 11391 (11700216) & 2010-03-03 & ACIS-S VFAINT & 24.80 & Balmaverde et al. (2012) \\ 
  285.0 & 13:21:17.868 & +42:35:14.91 & 0.0794 & 1.461 &   351.00 &  6911 (07701073) & 2006-03-18 & ACIS-S VFAINT & 39.62 & Hardcastle et al. (2006) \\
  286.0 & 13:31:08.292 & +30:30:32.95 & 0.8499 & 7.760 &  5341.81 & 15006 (14700660) & 2013-02-26 & ACIS-S FAINT  &  2.00 & Kuraszkiewicz et al. (2015) \\ 
  287.0 & 13:30:37.689 & +25:09:10.96 & 1.055  & 7.567 &  6995.09 &  3103 (03700781) & 2002-01-06 & ACIS-S VFAINT & 36.21 & Siemiginowska et al. (2008) \\ 
  288.0 & 13:38:49.9   & +38:51:09.5  & 0.246  & 3.777 &  1209.42 &  9257 (09700482) & 2008-04-13 & ACIS-S VFAINT & 39.64 & Hardcastle et al. (2009), Lal et al. (2010) \\ 
  289.0 & 13:45:26.251 & +49:46:32.47 & 0.9674 & 7.862 &  6643.17 & 15007 (14700660) & 2013-07-28 & ACIS-S FAINT  &  9.70 & Kuraszkiewicz et al. (2015) \\ 
  298.0 & 14:19:08.18  & +06:28:34.8  & 1.4381 & 8.401 & 10301.34 &  3104 (03700781) & 2002-03-01 & ACIS-S VFAINT & 17.88 & Siemiginowska et al. (2008) \\
  299.0 & 14:21:05.631 & +41:44:48.68 & 0.367  & 4.991 &  1923.58 & 12019 (10700678) & 2009-11-08 & ACIS-S VFAINT & 39.53 & \\
  309.1 & 14:59:07.58  & +71:40:19.9  & 0.905  & 7.717 &  5776.46 &  3105 (03700781) & 2002-01-28 & ACIS-S VFAINT & 16.95 & Belsole et al. (2006) \\ 
  310.0 & 15:04:57.12  & +26:00:58.5  & 0.0538 & 1.019 &   233.38 & 11845 (11700016) & 2010-04-09 & ACIS-S FAINT  & 57.58 & Kraft et al. (2012) \\ 
  318.0 & 15:20:05.484 & +20:16:05.75 & 1.574  & 8.435 & 11528.39 &  9256 (09700482) & 2008-05-05 & ACIS-S FAINT  &  9.78 & Wilkes et al. (2013) \\
  318.1 & 15:21:51.9   & +07:42:31.9  & 0.0453 & 0.867 &   195.26 &   900 (01800303) & 2000-04-03 & ACIS-I VFAINT & 57.32 & Mazzotta et al. (2002) \\ 
  324.0 & 15:49:48.811 & +21:25:38.34 & 1.2063 & 8.237 &  8270.74 &   326 (01600145) & 2000-06-25 & ACIS-S VFAINT & 42.18 & Boschin (2002) \\ 
  325.0 & 15:49:58.421 & +62:41:21.57 & 1.135  & 8.151 &  7664.22 &  6267 (05700521) & 2005-04-14 & ACIS-S VFAINT & 29.65 & Salvati et al. (2008) \\
  334.0 & 16:20:21.819 & +17:36:23.90 & 0.5551 & 6.335 &  3159.91 &  2097 (02700363) & 2001-08-22 & ACIS-S FAINT  & 32.47 & Hardcastle et al. (2004) \\
  336.0 & 16:24:39.090 & +23:45:12.23 & 0.9265 & 7.769 &  5948.01 & 15008 (14700660) & 2013-03-03 & ACIS-S FAINT  &  2.00 & Kuraszkiewicz et al. (2015) \\ 
  337.0 & 16:28:52.569 & +44:19:06.58 & 0.635  & 6.755 &  3724.80 & 15009 (14700660) & 2013-10-05 & ACIS-S FAINT  &  9.95 & Kuraszkiewicz et al. (2015) \\ 
  338.0 & 16:28:38.240 & +39:33:04.14 & 0.0304 & 0.592 &   129.53 & 10748 (10800906) & 2009-11-19 & ACIS-I VFAINT & 40.58 & Kirkpatrick et al. (2011), Nulsen et al. (2013) \\ 
  340.0 & 16:29:36.591 & +23:20:12.83 & 0.7754 & 7.331 &  4766.4  & 15010 (14700660) & 2013-10-20 & ACIS-S FAINT  &  9.95 & Kuraszkiewicz et al. (2015) \\
  343.0 & 16:34:33.809 & +62:45:35.89 & 0.988  & 7.905 &  6444.64 & 15011 (14700660) & 2013-04-28 & ACIS-S FAINT  &  9.94 & Kuraszkiewicz et al. (2015) \\ 
  343.1 & 16:38:28.203 & +62:34:44.29 & 0.75   & 7.240 &  4573.74 & 15012 (14700660) & 2013-02-25 & ACIS-S FAINT  &  9.94 & Kuraszkiewicz et al. (2015) \\ 
  352.0 & 17:10:44.138 & +46:01:28.47 & 0.8067 & 7.436 &  5006.3  & 15013 (14700660) & 2013-10-10 & ACIS-S FAINT  &  9.95 & Kuraszkiewicz et al. (2015) \\ 
  356.0 & 17:24:19.041 & +50:57:40.14 & 1.079  & 8.069 &  7194.59 &  9257 (09700482) & 2008-01-20 & ACIS-S FAINT  & 19.87 & Wilkes et al. (2013) \\ 
  368.0 & 18:05:06.3   & +11:01:32.0  & 1.131  & 8.146 &  7630.44 &  9258 (09700482) & 2008-06-01 & ACIS-S FAINT  & 19.91 & Wilkes et al. (2013) \\ 
  382.0 & 18:35:03.387 & +32:41:46.85 & 0.0579 & 1.092 &   251.98 &  6151 (05701042) & 2004-10-30 & ACIS-S FAINT  & 63.87 & Gliozzi et al. (2007) \\ 
  388.0 & 18:44:02.374 & +45:33:29.56 & 0.0917 & 1.663 &   408.83 &  5295 (05700009) & 2004-01-29 & ACIS-I VFAINT & 30.71 & Kraft et al. (2006) \\ 
  401.0 & 19:40:25.039 & +60:41:36.05 & 0.2011 & 3.236 &   962.99 &  4370 (03700685) & 2002-09-21 & ACIS-S FAINT  & 24.85 & Reynolds et al. (2005) \\ 
  427.1 & 21:04:06.966 & +76:33:10.28 & 0.572  & 6.430 &  3277.55 &  2194 (02700664) & 2002-01-27 & ACIS-S FAINT  & 39.45 & Hardcastle et al. (2004) \\ 
  432.0 & 21:22:46.327 & +17:04:37.96 & 1.785  & 8.424 & 13479.35 &  5624 (06700234) & 2005-01-07 & ACIS-S VFAINT & 19.78 & Erlund et al. (2006) \\ 
  433.0 & 21:23:44.582 & +25:04:27.63 & 0.1016 & 1.823 &   456.17 &  7881 (08700989) & 2007-08-28 & ACIS-S VFAINT & 37.17 & Miller \& Brandt (2009) \\ 
  437.0 & 21:47:25.265 & +15:20:32.03 & 1.48   & 8.415 & 10677.05 &  9259 (09700482) & 2008-01-07 & ACIS-S FAINT  & 19.88 & Wilkes et al. (2013) \\ 
  438.0 & 21:55:52.269 & +38:00:28.33 & 0.29   & 4.257 &  1460.96 & 12879 (12800244) & 2011-01-28 & ACIS-S VFAINT & 72.04 & Hardcastle et al. (2004) \\ 
  441.0 & 22:06:04.90  & +29:29:20.0  & 0.708  & 7.078 &  4259.10 & 15656 (14700660) & 2013-06-26 & ACIS-S FAINT  &  6.98 & Kuraszkiewicz et al. (2015) \\ 
  442.0 & 22:14:46.894 & +13:50:27.13 & 0.0263 & 0.515 &   111.71 &  6392 (06700371) & 2005-10-07 & ACIS-I VFAINT & 32.69 & Worrall et al. (2007), Hardcastle et al. (2007) \\ 
  449.0 & 22:31:20.582 & +39:21:29.53 & 0.0171 & 0.338 &    72.12 & 13123 (11800387) & 2010-09-20 & ACIS-S VFAINT & 59.92 & Lal et al. (2013) \\ 
  455.0 & 22:55:03.91  & +13:13:35.0  & 0.543  & 6.264 &  3249.99 & 15014 (14700660) & 2013-08-13 & ACIS-S FAINT  &  9.95 & Kuraszkiewicz et al. (2015) \\ 
  469.1 & 23:55:23.034 & +79:55:18.28 & 1.336  & 8.348 &  9396.53 &  9260 (09700482) & 2009-05-24 & ACIS-S FAINT  & 19.91 & Wilkes et al. (2013) \\ 
  470.0 & 23:58:35.910 & +44:04:45.51 & 1.653  & 8.439 & 12252.68 &  9261 (09700482) & 2008-03-03 & ACIS-S FAINT  & 19.91 & Wilkes et al. (2013) \\ 
\noalign{\smallskip}
\hline
\end{tabular}\\
Col. (1): The 3CR name.
Col. (2): Right ascension and Declination (equinox J2000) of the radio position used to perform the registration (see \S~\ref{sec:obs} for details). We reported here the original 3CR position \citep{spinrad85} of the sources for which the radio core was not clearly detected.
Col. (3): Redshift $z$. We also verified in the literature (e.g., NED and/or SIMBAD databases) if new $z$ values were reported after the release of the 3CR catalog.
Col. (4): The angular to linear scale factor in arcseconds. Cosmological parameters used to compute it are reported in \S~\ref{sec:intro}.
Col. (5): Luminosity Distance in Mpc. Cosmological parameters used to compute it are reported in \S~\ref{sec:intro}. 
Col. (6): \chn\ observation identification number. Proposal identification number is also reported in parenthesis.
Col. (7): \chn\ observation date.
Col. (8): Data mode indicates how the \chn\ ACIS detector was configured for the observation analyzed.
Col. (9): The total exposure.
Col. (10): The reference for the \chn\ observation.
\end{table*}

\begin{table}
\tiny
\caption{Summary of radio observations.}
\begin{center}
\label{tab:radio}
\begin{tabular}{|ccccc|}
\hline
\noalign{\smallskip}
Name & NRAO Project ID. & Freq & Time on source & HPBW \\
& & (GHz) & (sec) & (arcsec\,x\,arcsec) \\
\hline
\noalign{\smallskip}
3CR\,210 & AO230 & 1.42 & 1200 & 1.66 $\times$ 1.62 \\
3CR\,256 & AM224 & 4.76 & 180 & 1.72 $\times$ 1.35 \\
3CR\,267 & AL330 & 8.44 & 1620 & 0.81 $\times$ 0.74 \\
3CR\,277.1 & AV231 & 22.46 & 360 & 0.097 $\times$ 0.080 \\
3CR\,437 & AV164 & 4.86 & 1500 & 1.22 $\times$ 1.17 \\
3CR\,470 & AL330 & 8.46 & 1780 & 1.54 $\times$ 1.27 \\
\noalign{\smallskip}
\hline
\end{tabular}\\
\end{center}
Col. (1): The 3CR name.
Col. (2): The identification number of the observer program, as reported in the header of the raw u,v data downloaded from the VLA archive (see https://archive.nrao.edu/archive/nraodashelpj.html for more details).
Col. (3): The frequency at which the radio observations were performed.
Col. (4): The total exposure in seconds.
Col. (5): The half-power beam width (HPBW) of the reduced radio images.
\end{table}

\begin{table*} 
\tiny
\caption{X-ray emission from radio cores.}
\label{tab:cores1}
\begin{center}
\begin{tabular}{|rrrrrrrr|}
\hline
3CR  & Net Counts & Ext. Ratio & F$_{0.5-1~keV}^*$ & F$_{1-2~keV}^*$ & F$_{2-7~keV}^*$ & F$_{0.5-7~keV}^*$ & L$_X$ \\
name & &        & (cgs)                 & (cgs)           & (cgs)           & (cgs)             & (10$^{44}$erg~s$^{-1}$) \\
\hline 
\noalign{\smallskip}
2.0 & 839 (29) & 0.34 (0.02) & 77.79 (7.04) & 125.34 (6.16) & 291.05 (17.06) & 494.18 (19.45) & 31.07 (1.22)\\
13.0 &  14 (4) & 0.00 (-) & 0.79 (0.4) & 0.88 (0.4) & 3.02 (1.51) & 4.69 (1.61) & 0.57 (0.2)\\
14.0 & 228 (15) & 0.94 (0.09) & 59.38 (8.38) & 129.09 (12.48) & 314.41 (37.58) & 502.88 (40.47) & 75.48 (6.07)\\
22.0 &  64 (8) & 0.83 (0.14) & 0.0 (0.0) & 4.82 (1.64) & 95.56 (13.0) & 100.38 (13.11) & 4.9 (0.64)\\
35.0 &  12 (3) & 0.61 (0.24) & 0.0 (0.0) & 1.14 (0.47) & 2.93 (1.19) & 4.07 (1.28) & 0.0005 (0.0002)\\
40.0 &2443 (49) & 0.54 (0.02) & 80.69 (2.21) & 35.59 (1.45) & 95.95 (4.95) & 212.24 (5.61) & 0.00146 (0.00004)\\
43.0 & 162 (13) & 0.98 (0.11) & 42.9 (6.78) & 87.82 (10.28) & 216.29 (30.59) & 347.0 (32.97) & 52.17 (4.96)\\
48.0$^+$& 5814 (76) & 0.96 (0.02) & 699.0 (14.02) & 808.37 (17.05) & 1514.14 (46.08) & 3021.51 (51.09) & 15.0 (0.25)\\
49.0 & 156 (12) & 0.96 (0.11) & 0.31 (0.7) & 23.32 (3.33) & 162.79 (15.66) & 186.42 (16.03) & 3.28 (0.28)\\
65.0 & 196 (14) & 0.05 (0.02) & 0.89 (0.4) & 13.46 (1.6) & 77.54 (7.09) & 91.9 (7.28) & 7.91 (0.63)\\
68.1 & 41 (6) & 0.9 (0.19) & 4.85 (2.42) & 18.1 (4.84) & 115.99 (24.18) & 138.93 (24.78) & 13.6 (2.43)\\
68.2 &   9 (3) & 0.28 (0.12) & 0.21 (0.21) & 0.37 (0.26) & 5.08 (2.07) & 5.66 (2.1) & 1.01 (0.37)\\
75.0 & 219 (15) & 0.79 (0.07) & 23.62 (2.87) & 14.86 (1.73) & 57.96 (6.81) & 96.44 (7.59) & 0.0013 (0.0001)\\
78.0$^+$& 20856 (144) & 0.92 (0.01) & 432.13 (5.1) & 647.73 (6.62) & 1004.17 (15.55) & 2084.03 (17.65) & 0.0396 (0.0003)\\
88.0 & 659 (26) & 0.6 & 3.23 (0.61) (0.03) & 18.44 (1.24) & 109.08 (5.59) & 130.75 (5.76) & 0.0029 (0.0001)\\
98.0 & 1245 (35) & 0.93 (0.04) & 6.06 (1.24) & 9.53 (1.19) & 682.98 (20.18) & 698.57 (20.26) & 0.0153 (0.0004)\\
129.1 &  14 (4) & 0.22 (0.06) & 0.87 (0.87) & 4.45 (1.51) & 0.7 (2.27) & 6.03 (2.86) & 7.02e-5 (3.33e-5)\\
136.1 &   6 (2) & 0.17 (0.17) & 0.53 (0.53) & 0.54 (0.38) & 5.57 (3.22) & 6.64 (3.28) & 0.0007 (0.0003)\\
138.0$^+$& 385 (20) & 0.96 (0.07) & 116.59 (16.7) & 330.91 (25.32) & 1232.66 (96.62) & 1680.17 (101.27) & 48.56 (2.93)\\
147.0 & 150 (12) & 0.99 (0.11) & 44.12 (10.7) & 160.22 (17.8) & 341.21 (47.89) & 545.54 (52.2) & 6.99 (0.67)\\
172.0 &   26 (5) & 0.70 (0.18) & 0.0 (0.0) & 1.81 (0.91) & 43.38 (9.09) & 45.19 (9.13) & 0.51 (0.1)\\
175.0$^+$& 355 (19) & 0.95 (0.07) & 160.43 (19.35) & 308.97 (24.35) & 831.95 (75.32) & 1301.35 (81.49) & 38.73 (2.43)\\
175.1 &  86 (9) & 0.89 (0.13) & 4.91 (1.55) & 16.32 (2.56) & 43.87 (7.66) & 65.1 (8.23) & 3.04 (0.38)\\
181.0 & 166 (13)& 0.96 (0.10) & 55.43 (7.86) & 91.52 (10.51) & 181.96 (29.52) & 328.91 (32.3) & 42.39 (4.16)\\
184.0 &  38 (6) & 0.75 (0.16) & 0.72 (0.32) & 0.86 (0.38) & 21.92 (4.07) & 23.5 (4.1) & 1.33 (0.23)\\
190.0 & 165 (13) & 0.96 (0.10) & 49.91 (7.72) & 96.73 (10.62) & 150.5 (24.1) & 297.14 (27.44) & 26.74 (2.47)\\
191.0 & 715 (27) & 0.95 (0.05) & 32.66 (2.25) & 57.13 (3.17) & 121.57 (9.04) & 211.37 (9.84) & 64.62 (3.01)\\
192.0 & 46 (7) & 0.72 (0.15) & 1.77 (0.72) & 3.15 (1.13) & 52.72 (9.83) & 57.64 (9.92) & 0.0511 (0.0009)\\
196.0 & 87 (9) & 0.90 (0.13) & 10.4 (5.2) & 87.35 (13.75) & 297.91 (45.97) & 395.67 (48.26) & 16.1 (1.96)\\
200.0 & 202 (14) & 0.81 (0.08) & 11.55 (1.36) & 19.61 (2.1) & 32.8 (5.19) & 63.97 (5.76) & 0.54 (0.05)\\
204.0$^+$& 343 (19) & 0.96 (0.07) & 114.07 (11.21) & 190.04 (14.9) & 301.77 (35.56) & 605.89 (40.15) & 45.36 (3.01)\\
205.0$^+$& 969 (31&  0.95 (0.04) & 83.36 (5.42) & 160.57 (7.62) & 381.82 (22.71) & 625.75 (24.56) & 104.63 (4.11)\\
208.0 & 260 (16) & 0.98 (0.09) & 81.55 (9.68) & 135.04 (12.54) & 314.41 (36.8) & 531.0 (40.06) & 39.57 (2.99)\\
210.0 &  28 (5) & 0.31 (0.08) & 0.07 (0.16) & 1.7 (0.54) & 15.04 (3.67) & 16.82 (3.72) & 1.43 (0.32)\\
215.0$^+$& 11445 (107) & 0.96 (0.01) & 306.26 (5.02) & 480.6 (6.97) & 1140.72 (21.07) & 1927.58 (22.75) & 12.51 (0.15)\\
216.0$^+$& 244 (16) & 0.97 (0.09) & 134.19 (17.93) & 203.98 (19.72) & 553.56 (62.16) & 891.73 (67.63) & 18.94 (1.44)\\
220.1 & 1072 (33) & 0.79 (0.03) & 49.97 (2.44) & 70.94 (3.55) & 162.57 (10.71) & 283.48 (11.55) & 4.98 (0.2)\\
226.0 &  54 (7) & 0.83 (0.15) & 0.63 (0.63) & 4.33 (1.46) & 64.42 (9.97) & 69.38 (10.09) & 2.41 (0.35)\\
228.0 & 338 (18) & 0.93 (0.07) & 19.88 (1.9) & 34.71 (2.88) & 79.68 (8.92) & 134.28 (9.57) & 1.77 (0.13)\\
241.0 & 146 (12) & 1.00 (0.12) & 1.75 (0.58) & 13.25 (1.61) & 48.95 (5.89) & 63.96 (6.14) & 12.19 (1.17)\\
245.0 & 1835 (43) & 0.94 (0.03) & 154.08 (6.21) & 264.18 (9.48) & 608.4 (29.24) & 1026.67 (31.37) & 63.31 (1.93)\\
249.1$^+$& 4367 (66) & 0.96 (0.02) & 252.12 (8.42) & 322.33 (7.58) & 1041.37 (25.72) & 1615.82 (28.11) & 5.44 (0.09)\\
252.0 & 86 (9) & 0.64 (0.09) & 0.19 (0.19) & 4.25 (0.94) & 51.24 (6.52) & 55.67 (6.59) & 4.08 (0.48)\\
\noalign{\smallskip}
\hline
\end{tabular}\\
\end{center}
Col. (1): The 3CR name.
Col. (2): The net counts. The 1 $\sigma$ uncertainties, reported in parenthesis, are computed assuming a Poisson distribution.
Col. (3): The Ext. Ratio defined as the ratio of the net counts in the r\,=\,2\arcsec\ circle to the net counts in the
r\,=\,10\arcsec\ circular region surrounding the core of each 3CR source. The 1 $\sigma$ uncertainties, reported in parenthesis, are computed assuming a Poisson distribution.
Col. (4): Measured X-ray flux between 0.5 and 1 keV.
Col. (5): Measured X-ray flux between 1 and 2 keV.
Col. (6): Measured X-ray flux between 2 and 7 keV.
Col. (7): Measured X-ray flux between 0.5 and 7 keV.
Col. (8): X-ray luminosity in the range 0.5 to 7 keV with the 1$\sigma$ uncertainties given in parenthesis.\\
Note:\\
($^*$) Fluxes are given in units of 10$^{-15}$erg~cm$^{-2}$s$^{-1}$ and 1$\sigma$ uncertainties are given in parenthesis. 
The uncertainties on the flux measurements were computed as described in \S~\ref{sec:xray}\\
($^+$) Sources having count rates above the threshold of 0.1 counts per frame 
for which the X-ray flux measurements is affected by pileup \citep[see][and references therein for additional details]{massaro13}.
\end{table*}

\begin{table*} 
\tiny
\caption{X-ray emission from radio cores.}
\label{tab:cores2}
\begin{center}
\begin{tabular}{|rrrrrrrr|}
\hline
3CR  & Net Counts & Ext. Ratio & F$_{0.5-1~keV}^*$ & F$_{1-2~keV}^*$ & F$_{2-7~keV}^*$ & F$_{0.5-7~keV}^*$ & L$_X$ \\
name & &        & (cgs)                 & (cgs)           & (cgs)           & (cgs)             & (10$^{44}$erg~s$^{-1}$) \\
\hline 
\noalign{\smallskip}
263.1 & 430 (21) & 0.96 (0.07) & 52.7 (5.0) & 70.61 (5.15) & 180.43 (15.76) & 303.74 (17.32) & 10.77 (0.61)\\
266.0 &  19 (4) &1.07 (0.42) & 0.5 (0.29) & 0.81 (0.41) & 9.29 (2.68) & 10.61 (2.73) & 1.12 (0.29)\\
267.0 & 166 (13) & 0.89 (0.10) & 0.88 (0.4) & 10.0 (1.41) & 81.65 (7.83) & 92.53 (7.97) & 7.37 (0.63)\\
268.1 &  46 (7) & 0.94 (0.20) & 0.4 (0.4) & 1.21 (0.88) & 60.37 (9.31) & 61.98 (9.36) & 3.32 (0.5)\\
268.3 & 398 (20) & 0.98 (0.07) & 1.3 (0.3) & 4.57 (0.66) & 117.52 (6.54) & 123.4 (6.58) & 0.63 (0.03)\\
268.4 & 282 (17) & 0.98 (0.08)& 78.15 (9.55) & 145.82 (12.99) & 375.22 (40.28) & 599.19 (43.39) & 79.76 (5.78)\\
270.1$^+$& 691 (26) & 0.94 (0.05) & 69.18 (4.89) & 120.54 (6.72) & 219.62 (17.11) & 409.34 (19.02) & 66.79 (3.1)\\
277.1 & 2287 (48) & 0.95 (0.03) & 167.79 (5.7) & 225.01 (7.37) & 468.66 (21.27) & 861.45 (23.22) & 3.1 (0.08)\\
277.3 & 229 (15) & 0.81 (0.07) & 1.38 (0.5) & 4.28 (0.83) & 140.84 (10.18) & 146.5 (10.22) & 0.028 (0.002)\\
285.0 & 457 (21) & 0.87 (0.06) & 0.34 (0.2) & 1.44 (0.38) & 216.95 (10.38) & 218.73 (10.39) & 0.036 (0.002)\\
286.0 & 117 (11) & 0.96 (0.12) & 87.64 (14.08) & 100.04 (13.74) & 158.45 (31.69) & 346.14 (37.3) & 13.21 (1.42)\\
287.0 & 3424 (59) & 0.97 (0.02) & 95.29 (2.62) & 129.05 (3.45) & 245.11 (9.29) & 469.45 (10.25) & 30.81 (0.67)\\
289.0 &  52 (7) & 0.75 (0.13) & 0.0 (0.0) & 3.01 (1.37) & 78.49 (11.63) & 81.5 (11.71) & 4.3 (0.62)\\
298.0$^+$& 9993 (100) & 0.97 (0.01) & 493.95 (8.59) & 821.4 (12.42) & 1660.05 (34.61) & 2975.41 (37.76) & 424.2 (5.38)\\
299.0 &  81 (9) & 0.77 (0.12) & 0.9 (0.28) & 0.53 (0.22) & 30.86 (3.81) & 32.29 (3.82) & 0.16 (0.02)\\
309.1$^+$& 5254 (72) & 0.97 (0.02) & 259.9 (6.33) & 423.95 (9.19) & 1145.75 (30.3) & 1829.6 (32.29) & 81.67 (1.44)\\
318.0 & 256 (16) & 0.96 (0.08) & 23.6 (2.87) & 43.99 (4.0) & 89.4 (11.0) & 156.99 (12.05) & 4.43 (0.34)\\
318.1 & 106 (10) & 0.071 (0.004)& 2.79 (0.99) & 4.82 (1.09) & 3.97 (2.37) & 11.57 (2.79) & 0.0006 (0.0002)\\
324.0 &  40 (6) & 0.61 (0.14) & 0.64 (0.18) & 0.92 (0.27) & 5.48 (1.49) & 7.05 (1.52) & 0.65 (0.14)\\
325.0 & 365 (19) & 0.86 (0.06) & 2.6 (0.57) & 19.23 (1.58) & 93.96 (6.8) & 115.79 (7.01) & 4.56 (0.28)\\
334.0$^+$& 7178 (85) & 0.96 (0.02) & 203.45 (3.98) & 292.96 (5.48) & 684.21 (16.74) & 1180.62 (18.06) & 15.81 (0.24)\\
336.0$^+$& 191 (14) & 0.95 (0.10) & 98.63 (15.04) & 184.79 (19.01) & 343.29 (47.61) & 626.71 (53.42) & 29.78 (2.54)\\
337.0 &   9 (3) & 0.53 (0.23) & 0.0 (0.0) & 1.1 (0.79) & 11.33 (4.28) & 12.43 (4.35) & 0.23 (0.08)\\
338.0 & 246 (16) & 0.092 (0.005) & 16.48 (2.67) & 13.33 (2.08) & 12.17 (3.91) & 41.97 (5.17) & 0.0009 (0.0001)\\
340.0 &  86 (9) & 0.92 (0.14) & 1.46 (0.86) & 11.56 (2.32) & 84.78 (11.25) & 97.8 (11.52) & 2.98 (0.35)\\
343.0 &  18 (4) & 0.76 (0.25) & 2.88 (1.18) & 2.46 (1.02) & 6.61 (2.96) & 11.95 (3.34) & 0.67 (0.19)\\
343.1 & 47 (7) & 1.04 (0.22) & 3.02 (1.23) & 9.88 (2.06) & 25.2 (6.15) & 38.09 (6.6) & 1.07 (0.19)\\
352.0 & 129 (11) & 0.88 (0.11) & 2.57 (1.31) & 22.66 (3.21) & 95.99 (11.47) & 121.22 (11.98) & 4.05 (0.4)\\
356.0 &  24 (5) & 0.38 (0.09) & 0.35 (0.33) & 0.73 (0.43) & 13.24 (3.12) & 14.32 (3.17) & 0.99 (0.22)\\
382.0$^+$& 14052 (119) & 0.86 (0.01) & 72.82 (1.91) & 174.65 (3.14) & 2363.24 (24.23) & 2610.71 (24.5) & 0.215 (0.002)\\
388.0 & 271 (16) & 0.28 (0.02) & 20.53 (2.4) & 19.15 (1.82) & 19.28 (3.63) & 58.96 (4.71) & 0.013 (0.001)\\
401.0 & 229 (15) & 0.34 (0.02) & 9.48 (1.12) & 12.35 (1.44) & 26.96 (3.96) & 48.79 (4.36) & 0.06 (0.01)\\
427.1 &  18 (4) & 0.22 (0.05) & 0.24 (0.16) & 0.24 (0.24) & 4.85 (1.56) & 5.33 (1.59) & 0.08 (0.02)\\
432.0 & 730 (27) & 0.93 (0.05) & 34.32 (2.33) & 57.96 (3.23) & 120.42 (8.81) & 212.7 (9.67) & 53.29 (2.42)\\
433.0$^+$& 2724 (52) & 0.92 (0.02) & 2.69 (0.55) & 13.24 (1.25) & 1139.84 (22.48) & 1155.78 (22.52) & 0.32 (0.01)\\
437.0 &   7 (3) & 0.43 (0.19) & 0.2 (0.2) & 0.37 (0.26) & 4.1 (2.09) & 4.67 (2.12) & 0.71 (0.32)\\
438.0 & 162 (13) & 0.1 (0.01)& 1.2 (0.38) & 3.62 (0.62) & 14.83 (2.22) & 19.65 (2.34) & 0.06 (0.01)\\
442.0 & 181 (13) & 0.58 (0.06) & 3.08 (0.99) & 13.86 (1.53) & 41.91 (4.58) & 58.85 (4.93) & 0.00096 (8e-5)\\
449.0 & 558 (24) & 0.54 (0.03) & 12.81 (0.97) & 13.41 (0.95) & 31.81 (2.77) & 58.02 (3.08) & 0.0004 (2e-5)\\
455.0 & 150 (12) & 0.96 (0.11) & 13.61 (2.62) & 29.18 (3.39) & 64.09 (9.16) & 106.88 (10.11) & 1.35 (0.13)\\
469.1 &  77 (9) & 0.72 (0.11) & 0.32 (0.23) & 2.66 (0.78) & 47.29 (6.02) & 50.27 (6.07) & 5.95 (0.72)\\
470.0 & 54 (7) & 0.76 (0.14 )& 0.0 (0.0) & 1.46 (0.61) & 35.08 (5.06) & 36.54 (5.1) & 7.36 (1.03)\\
\noalign{\smallskip}
\hline
\end{tabular}\\
\end{center}
Col. (1): The 3CR name.
Col. (2): The net counts. The 1 $\sigma$ uncertainties, reported in parenthesis, are computed assuming a Poisson distribution.
Col. (3): The Ext. Ratio defined as the ratio of the net counts in the r\,=\,2\arcsec\ circle to the net counts in the
r\,=\,10\arcsec\ circular region surrounding the core of each 3CR source. The 1 $\sigma$ uncertainties, reported in parenthesis, are computed assuming a Poisson distribution.
Col. (4): Measured X-ray flux between 0.5 and 1 keV.
Col. (5): Measured X-ray flux between 1 and 2 keV.
Col. (6): Measured X-ray flux between 2 and 7 keV.
Col. (7): Measured X-ray flux between 0.5 and 7 keV.
Col. (8): X-ray luminosity in the range 0.5 to 7 keV with the 1$\sigma$ uncertainties given in parenthesis.\\
Note:\\
($^*$) Fluxes are given in units of 10$^{-15}$erg~cm$^{-2}$s$^{-1}$ and 1$\sigma$ uncertainties are given in parenthesis.
The uncertainties on the flux measurements were computed as described in \S~\ref{sec:xray}\\
($^+$) Sources having count rates above the threshold of 0.1 counts per frame 
for which the X-ray flux measurements is affected by pileup \citep[see][and references therein for additional details]{massaro13}.
\end{table*}

\begin{table*} 
\tiny
\caption{X-ray emission from radio extended structures (i.e., knots and hotspots).}
\label{tab:features}
\begin{center}
\begin{tabular}{|rrrrrrrrr|}
\hline
3CR  & Component & class & Counts & F$_{0.5-1~keV}^*$ & F$_{1-2~keV}^*$ & F$_{2-7~keV}^*$ & F$_{0.5-7~keV}^*$ & L$_X$ \\
name &        & & & (cgs)                 & (cgs)           & (cgs)           & (cgs)             & (10$^{42}$erg~s$^{-1}$) \\
\hline 
\noalign{\smallskip}
13.0 & n16.5 & h & 5 (0.375) & 0.58 (0.34) & 0.16 (0.16) & 0.76 (0.76) & 1.5 (0.85) & 18.27 (10.35)\\
65.0 & e6.6 & h & 4 (0.375) & 0.23 (0.23) & 0.0 (0.0) & 1.01 (1.01) & 1.24 (1.04) & 10.68 (8.95)\\
65.0 & w6.7 & h & 7 (0.25) & 0.46 (0.27) & 0.4 (0.29) & 0.72 (0.72) & 1.58 (0.82) & 13.6 (7.06)\\
68.2 & n11.5 & h & 5 (0.25) & 0.0 (0.0) & 1.02 (0.46) & 0.0 (0.0) & 1.02 (0.46) & 18.21 (8.21)\\
78.0 & e1.6 & k & 1001 (406) & 15.58 (1.29) & 22.46 (1.61) & 39.19 (3.84) & 77.22 (4.36) & 0.15 (0.01)\\
88.0 & e109 & k & 33 (6.75) & 0.71 (0.26) & 0.77 (0.25) & 0.72 (1.02) & 2.2 (1.08) & 0.005 (0.002)\\
181.0 & e4.5 & h & 3 (0.125) & 1.21 (1.21) & 0.94 (0.94) & 2.7 (2.7) & 4.86 (3.11) & 62.64 (40.08)\\
191.0 & s1.9$^+$ & k & 18 (0.125) & 1.01 (0.42) & 0.86 (0.44) & 3.19 (1.6) & 5.07 (1.71) & 154.99 (52.28)\\
200.0 & s9.3$^+$ & k & 6 (0.125) & 0.0 (0.0) & 0.7 (0.41) & 1.49 (1.49) & 2.19 (1.54) & 1.84 (1.3)\\
210.0 & s7.6 & h & 5 (0.125) & 0.28 (0.2) & 0.28 (0.28) & 0.65 (0.65) & 1.21 (0.74) & 10.26 (6.28)\\
215.0 & e2.6$^+$ & k & 26 (0.25) & 0.31 (0.28) & 0.14 (0.36) & 1.06 (1.12) & 1.5 (1.21) & 0.97 (0.79)\\
228.0 & n24.8 & h & 6 (0.125) & 0.0 (0.0) & 1.03 (0.46) & 0.59 (0.59) & 1.62 (0.75) & 2.14 (0.99)\\
228.0 & s21.4 & h & 16 (0.125) & 1.35 (0.48) & 2.25 (0.8) & 0.0 (0.0) & 3.6 (0.93) & 4.76 (1.23)\\
245.0 & w1.5$^+$ & k & 26 (0.125) & 1.98 (0.76) & 0.76 (1.07) & 1.16 (2.32) & 3.9 (2.66) & 24.05 (16.4)\\
268.1 & w25 & h & 25 (0.125) & 0.71 (0.5) & 5.95 (1.49) & 11.08 (4.19) & 17.74 (4.47) & 95.13 (23.97)\\
299.0 & e2.7 & h & 22 (0.375) & 0.94 (0.28) & 1.0 (0.32) & 0.45 (0.45) & 2.39 (0.62) & 1.19 (0.31)\\
324.0 & e5.8 & h & 9 (0.5) & 0.18 (0.1) & 0.23 (0.12) & 0.42 (0.42) & 0.84 (0.45) & 7.7 (4.13)\\
325.0 & e6.8 & h & 10 (0.375) & 0.34 (0.19) & 0.34 (0.25) & 0.95 (0.68) & 1.63 (0.75) & 6.43 (2.96)\\
325.0 & w9.2 & h & 7 (0.125) & 0.1 (0.1) & 0.21 (0.21) & 1.73 (0.86) & 2.03 (0.89) & 8.0 (3.51)\\
334.0 & s2.7$^+$ & k & 30 (0.625) & 1.0 (0.32) & 0.94 (0.4) & 0.25 (0.81) & 2.19 (0.95) & 2.93 (1.27)\\
334.0 & s17.5$^+$ & k & 26 (5.75) & 0.76 (0.27) & 1.1 (0.37) & 1.99 (1.82) & 3.85 (1.87) & 5.15 (2.5)\\
437.0 & n19 & h & 12 (0.625) & 0.48 (0.28) & 0.74 (0.37) & 3.32 (1.66) & 4.54 (1.72) & 69.43 (26.3)\\
437.0 & s17 & h & 7 (0.5) & 0.32 (0.23) & 0.22 (0.22) & 1.28 (1.19) & 1.83 (1.24) & 27.98 (18.96)\\
470.0 & n14.4 & h & 1 (0.75) & 0.0 (0.0) & 0.0 (0.0) & 0.29 (0.66) & 0.29 (0.66) & 5.84 (13.29)\\
470.0 & s9.4 & h & 10 (0.625) & 0.76 (0.38) & 0.33 (0.24) & 2.0 (1.42) & 3.1 (1.49) & 62.43 (30.01)\\
\noalign{\smallskip}
\hline
\end{tabular}\\
\end{center}
Col. (1): The 3CR name.
Col. (2): The component name chosen according to the definition reported in \S~\ref{sec:xray}.
Col. (3): The component class: ``h'' = hotspot - ``k'' = knot.
Col. (4): The number of counts column gives the total counts in the photometric circle 
together with the average of the 8 background regions, in parentheses; both for the 0.5 to 7 keV band.
Col. (5): Measured X-ray flux between 0.5 and 1 keV.
Col. (6): Measured X-ray flux between 1 and 2 keV.
Col. (7): Measured X-ray flux between 2 and 7 keV.
Col. (8): Measured X-ray flux between 0.5 and 7 keV.
Col. (9): X-ray luminosity in the range 0.5 to 7 keV with the 1$\sigma$ uncertainties given in parenthesis.\\
Note:\\
($^+$) Source components for which the X-ray emission was already reported in the literature.\\
($^*$) Fluxes are given in units of 10$^{-15}$erg~cm$^{-2}$s$^{-1}$ and 1$\sigma$ uncertainties are given in parenthesis.
The uncertainties on the flux measurements were computed as described in \S~\ref{sec:xray}\\
\end{table*}

\begin{table*} 
\tiny
\caption{X-ray galaxy clusters.}
\label{tab:clusters}
\begin{center}
\begin{tabular}{|llrrrr|}
\hline
3CR  & Other    & $z$ & R           & R        & Total  \\
name &  name  &        & (arcsec) & (kpc)  &  Counts                  \\
\hline 
\noalign{\smallskip}
28.0      & Abell 115				& 0.195		& 200		& 632	& 31450 \\
40.0      & Abell 194				& 0.0181	& 170		&  60	& 23855 \\
75.0      & Abell 400				& 0.023		& 500		& 228	& 57400 \\
88.0      & 1RXS J032755.0+023403		& 0.0302	& 120		&  70	&  6610 \\
220.1   & 1RXS J093245.5+790636		& 0.61	 	& 25		& 166	&  1722 \\
288.0     & 1RXS J133849.3+385110		& 0.246	 	& 60		& 680	&  5324 \\
310.0     & SDSS J150457.12+260058.4		& 0.0538	& 180		& 183	& 28309 \\
318.1   & Abell 2063B				& 0.0453	& 500		& 433	&272770 \\
338.0     & Abell 2199				& 0.03035	& 500		& 296	&504360 \\
388.0     & 1RXS J184402.1+453332		& 0.0917	& 240		& 400	& 15416 \\
401.0     & 1RXS J194024.4+604136		& 0.055	 	& 90		& 292	&  3200 \\
427.1   &					& 0.572	 	& 40		& 257	&   467 \\
438.0     & 1RXS J215553.4+380021		& 0.290		& 210		& 894	& 66288 \\
442.0     & 1RXS J221451.0+135040		& 0.0263	& 300		& 155	& 15506 \\
449.0     & 					& 0.017		& 240		&  81	& 42378 \\
\noalign{\smallskip}
\hline
\end{tabular}\\
\end{center}
Col. (1): The 3CR name.
Col. (2): Alternative name.
Col. (3): The source redshift.
Col. (4): Radius in arcseconds.
Col. (5): Radius in kpc.
Col. (6): The net counts.
\end{table*}

\clearpage
\appendix

\section{A: Images of the sources}
\label{sec:images}

For all the 93 3CR sources in our selected sample, radio morphologies are shown here 
as contours superposed on the re-gridded/smoothed X-ray events files.  
The full width half maximum (FWHM) of the Gaussian
smoothing function and the binning factor are given in Table~\ref{tab:captions1} and ~\ref{tab:captions2}.
X-ray event files were limited to the 0.5 to 7 keV band and rebinned
to change the pixel size with a binning factor 'f' (e.g. f=1/4
produces pixels 4 times smaller than the native ACIS pixel of 0.492\arcsec).  
The labels on the color bar for each X-ray map are in units of
counts/pixel. Also included in this table are the radio
brightness of the lowest contour, the factor (usually 2 or 4) by which
each subsequent contour exceeds the previous one, the frequency of the
radio map, and the FWHM of the clean beam.
The primary reason figures appear so different from
each other is the wide range in angular size of the radio sources.

\begin{table*} 
\tiny
\caption{Captions}
\label{tab:captions1}
\begin{tabular}{|lrrrrrrr|}
\hline
3CR  & Binning Factor & FWHM-smoothing & Low-contour-level & Factor   & Radio freq. & HPBW & NRAO Project Code \\
name & (X-rays) & (X-rays)-(arcsec)      & (mJy/beam)         & increase & (GHz) & (arcsec\,x\,arcsec)  &         \\ 
\hline 
\noalign{\smallskip}
  2$^{(1)}$   & 1/4 & 0.72 & 6.4 & 4 & 1.5 & 1.2x1.4 & AH0171\,--\,(NVAS)\\
  13.0        & 1/8 & 0.51 & 1.0 & 2 & 4.9 & 0.37 & AC0200\,--\,(NVAS)\\
  14.0        & 1/8 & 0.51 & 3.0 & 4 & 8.5 & 0.23 & AL0280\,--\,(NVAS)\\
  22.0        & 1/8 & 0.51 & 0.125 & 4 & 8.5 & 0.25 & AP380\,--\,(MJH)\\
  28.0        & 1/8 & 0.65 & 0.25 & 4 & 1.4 & 1.10 & AL272\,--\,(NED)\\
  35.0        & 1   & 4.0  & 4.0 & 2 & 1.5 & 17x14 & AW0087\,--\,(NVAS)\\
  40.0        & 2   & 8.0  & 4.0 & 4 & 1.6 & 23x12 & AB0022\,--\,(NVAS)\\
  43.0        & 1/8 & 0.51 & 6.0 & 4 & 8.3 & 0.23 & AJ0206\,--\,(NVAS)\\
  48.0        & 1/8 & 0.36 & 12.5 & 4 & 4.8 & 0.59x0.47 & AW0227\,--\,(NVAS)\\
  49.0        & 1/8 & 0.36 & 4.0 & 4 & 4.9 & 0.41 & NEFF\,--\,(NED)\\
  65.0        & 1/8 & 0.51 & 4.0 & 2 & 1.5 & 1.28x1.13 & PERL\,--\,(NVAS)\\
  68.1        & 1/4 & 1.0  & 2.0 & 2 & 1.4 & 1.46x1.33 & AW0482\,--\,(NVAS)\\
  68.2        & 1/4 & 0.72 & 1.0 & 2 & 4.9 & 0.53x0.39 & AV0164\,--\,(NVAS)\\
  75.0        & 1   & 1.7  & 0.25 & 4 & 4.6 & 4.6x3.8 & AE0061\,--\,(NVAS)\\
  78.0$^{(2)}$& 1/8 & 0.51 & 2.0 & 2 & 1.5 & 4.3x3.9 & AB0376\,--\,(NVAS)\\
  88.0        & 1/2 & 1.4  & 1.0 & 2 & 4.9 & 4.4x4.2 & AP0077\,--\,(NVAS)\\
  98.0        & 1   & 4.0  & 0.25 & 2 & 8.3 & 2.0 & PERL\,--\,(NED)\\
  99.0        & 1/4 & --  & -- & -- & 4.8 & 0.45x0.39 & AS302\,--\,(NVAS)\\
  129.1       & 1/4 & 1.0  & 0.125 & 2 & 4.8 & 1.25 & AT229\,--\,(NVAS)\\
  136.1       & 1   & 3.5  & 2.0 & 2 & 1.6 & 3.3 & POOL\,--\,(NVAS)\\
  138.0       & 1/8 & 0.51 & 1.0 & 4 & 4.9 & 0.42 & AL0142\,--\,(NVAS)\\
  147.0       & 1/8 & 0.36 & 19.2 & 4 & 8.4 & 0.27 & AK403\,--\,(CCC)\\
  172.0       & 1/4 & 0.72 & 0.125 & 4 & 8.5 & 0.90 & AP361\,--\,(MJH)\\
  175.0       & 1/4 & 0.72 & 0.5 & 4 & 8.5 & 0.78x0.61 & AH0452\,--\,(NVAS)\\
  175.1       & 1/8 & 0.36 & 1.2 & 4 & 4.9 & 0.35 & AP380\,--\,(MJH)\\
  181.0       & 1/8 & 0.36 & 1.0 & 4 & 4.9 & 0.37 & AH0552\,--\,(NVAS)\\
  184.0       & 1/8 & 0.36 & 16.0 & 4 & 8.5 & 0.36x0.20 & AK0403\,--\,(NVAS)\\
  190.0       & 1/8 & 0.36 & 4.0 & 2 & 8.5 & 0.20 & AO0105\,--\,(NVAS)\\
  191.0       & 1/8 & 0.36 & 0.3 & 4 & 4.7 & 0.30 & AK180\,--\,(CCC)\\
  192.0       & 1/2 & 1.4  & 0.125 & 4 & 8.2 & 0.80 & PERL\,--\,(NED)\\
  196.0       & 1/8 & 0.36 & 0.5 & 4 & 4.9 & 0.35 & AB516\,--\,(MJH)\\
  200.0       & 1/8 & 0.65 & 0.125 & 4 & 8.5 & 0.25 & AP331\,--\,(MJH)\\
  204.0       & 1/8 & 0.51 & 0.5 & 4 & 8.3 & 0.78x0.65 & AW0249\,--\,(NVAS)\\
  205.0       & 1/8 & 0.36 & 2.0 & 4 & 8.3 & 0.22 & AW0330\,--\,(NVAS)\\
  208.0       & 1/8 & 0.36 & 0.125 & 4 & 8.4 & 0.25 & AL280\,--\,(CCC)\\
  210.0       & 1/4 & 1.3  & 6.4 & 4 & 1.4 & 1.6 & AO230\,--\,(TW)\\
  215.0$^{(3)}$ & 1/8 & 0.36 & 0.1 & 4 & 4.9 & 0.37 & BRID\,--\,(MJH)\\
  216.0       & 1/8 & 0.36 & 4.8 & 4 & 8.2 & 0.25 & AG357\,--\,(MJH)\\
  220.1       & 1/8 & 0.36 & 0.1 & 4 & 8.4 & 0.25 & AP380\,--\,(MJH)\\
  220.3$^{(4)}$ & 1/8 & 0.51 & 1.0 & 2 & 8.4 & 0.70x0.43 & AM0384\,--\,(NVAS)\\
  226.0       & 1/8 & 0.51 & 0.125 & 4 & 8.5 & 0.20 & AP380\,--\,(MJH)\\
  228.0       & 1/8 & 0.51 & 0.125 & 4 & 8.5 & 0.23 & AP331\,--\,(MJH)\\
  241.0       & 1/8 & 0.36 & 1.2 & 4 & 8.4 & 0.20 & AA0149\,--\,(NVAS)\\
  245.0       & 1/8 & 0.36 & 0.5 & 4 & 4.9 & 0.25 & AB244\,--\,(CCC)\\
  249.1       & 1/8 & 0.36 & 0.1 & 4 & 4.9 & 0.35 & BRID\,--\,(MJH)\\
  252.0       & 1/4 & 1.0 & 0.5 & 4 & 4.9 & 1.0 & AF0213\,--\,(NVAS)\\
  256.0       & 1/4 & 1.0 & 0.4 & 4 & 4.8 & 1.7x1.4 & AM244\,--\,(TW)\\
\noalign{\smallskip}
\hline
\end{tabular}\\
Col. (1): The 3CR name.
Col. (2): The binning factor of the X-ray image (see \S~\ref{sec:xray} for more details).
Col. (3): the full width half maximum (FWHM) of the smoothing kernel chosen for the X-ray image.
Col. (4): The value of the lowest contour level of the radio map overlaid to the X-ray image.
Col. (5): The factor increase of the radio contours.
Col. (6): The radio frequency of the radio map used for the comparison with the X-ray image.
Col. (7): The half-power beam width (HPBW) of the reduced radio images. Single numbers are reported for circular beam.
Col. (8): The identification number of the observer program, as reported in the header of the raw u,v data downloaded from the VLA archive (see https://archive.nrao.edu/archive/nraodashelpj.html for more details).\\
Notes:\\
1. 3CR 2 is 7.9\arcmin\ off axis, so we did not register the X-ray image. 
The source appears to be extended in the X-rays ($\sim$8\arcsec) but
since it is so far off axis, the apparent size is consistent with the \chn\ point spread function.\\
2. The white contours at the nucleus are from a Merlin observation, performed at 1.4 GHz on May 5$^{th}$ 1998, showing
the small scale jet.  The X-ray images comes from a 1/8th subarray observation
and the width of the subarray is smaller than the size of the radio source.
The readout streak is evident and lies, unfortunately, along the direction of the 
jet and the primary axis of the radio emission.\\
3. The prominent readout streak goes right through the jet segment superposed on the E lobe.\\
4. This is a radio galaxy lying at z=0.685 which is lensing a submillimeter galaxy at z=2.221 \citep{haas14}.\\
\end{table*} 

\begin{table*} 
\tiny
\caption{Captions}
\label{tab:captions2}
\begin{tabular}{|lrrrrrrr|}
\hline
3CR  & Binning Factor & FWHM-smoothing & Low-contour-level & Factor   & Radio freq. & HPBW & NRAO Project Code \\
name & (X-rays) & (X-rays)-(arcsec)      & (mJy/beam)         & increase & (GHz) & (arcsec\,x\,arcsec)  &         \\ 
\hline 
\noalign{\smallskip}
  263.1       & 1/8 & 0.36 & 0.25 & 4 & 4.9 & 0.35 & DREH\,--\,(MJH)\\
  266.0       & 1/8 & 0.51 & 0.4 & 4 & 8.4 & 0.30 & AK403\,--\,(CCC)\\
  267$^{(5)}$ & 1/8 & 0.36 & 0.125 & 4 & 8.4 & 0.85x0.74 & AL330\,--\,(TW)\\
  268.1       & 1/8 & 0.51 & 0.25 & 4 & 8.5 & 0.25 & AP380\,--\,(MJH)\\
  268.3       & 1/16 & 0.33 & 1.2 & 4 & 5.0 & 0.06 & MERLIN2\,--\,(MJH)\\
  268.4       & 1/8 & 0.36 & 2.0 & 4 & 8.3 & 0.72x0.58 & AW0249\,--\,(NVAS)\\
  270.1       & 1/8 & 0.51 & 4.8 & 4 & 4.9 & 0.36 & AB0522\,--\,(NVAS)\\
  277.1       & 1/8 & 0.36 & 1.0 & 4 & 22.5 & 0.09 & AV231\,--\,(TW)\\
  277.3       & 1/8 & 0.36 & 1.0 & 2 & 4.9 & 0.37 & CORD\,--\,(NVAS)\\
  285.0       & 1/4 & 0.72 & 0.3 & 4 & 1.5 & 1.2 & AV0127\,--\,(NVAS)\\
  286.0       & 1/8 & 0.36 & 4.8 & 4 & 8.0 & 0.28x0.23 & AG0357\,--\,(NVAS)\\
  287.0       & 1/16 & 0.18 & 10.0 & 4 & 8.5 & 0.24 & AK0276\,--\,(NVAS)\\
  288.0       & 1/4 & 1.3 & 0.5 & 4 & 4.9 & 0.6 & ED\,--\,(NED)\\
  289.0       & 1/8 & 0.51 & 0.4 & 4 & 5.0 & 0.06 & MERLIN2\,--\,(MJH)\\
  298.0       & 1/16 & 0.18 & 10.0 & 4 & 8.3 & 0.25 & AJ0206\,--\,(NVAS)\\
  299.0       & 1/8 & 0.36 & 1.0 & 4 & 1.5 & 0.13 & MERLIN2\,--\,(NED)\\
  309.1       & 1/8 & 0.36 & 4.0 & 4 & 14.9 & 0.17x0.11 & TESTT\,--\,(NVAS)\\
  310.0       & 1   & 7.5 & 10.0 & 2 & 1.5 & 15x12 & AB0182\,--\,(NVAS)\\
  318.0       & 1/8 & 0.51 & 20.0 & 2 & 8.5 & 0.22 & AA0149\,--\,(NVAS)\\
  318.1       & 1/2 & 2.0 & 0.3 & 4 & 1.4 & 4.7x4.4 & FOMA\,--\,(NVAS)\\
  324.0       & 1/8 & 0.36 & 0.125 & 4 & 4.9 & 0.38 & AF186\,--\,(CCC)\\
  325.0       & 1/8 & 0.51 & 0.5 & 4 & 4.9 & 0.35 & AF213\,--\,(MJH)\\
  334.0       & 1/8 & 0.51 & 0.125 & 4 & 4.9 & 0.35 & BRID\,--\,(MJH)\\
  336.0       & 1/4 & 0.72 & 0.1 & 4 & 4.9 & 0.35 & AB454\,--\,(MJH)\\
  337.0       & 1/4 & 1.0 & 0.25 & 4 & 4.9 & 0.40 & AP114\,--\,(MJH)\\
  338.0       & 1/4 & 0.72 & 0.1 & 2 & 4.9 & 1.0 & AG269\,--\,(NED)\\
  340.0       & 1/4 & 1.0 & 0.125 & 4 & 4.9 & 0.40 & AP380\,--\,(MJH)\\
  343.0       & 1/4 & 0.72 & 19.2 & 4 & 4.9 & 0.42 & AB0922\,--\,(NVAS)\\
  343.1       & 1/8 & 0.51 & 32.0 & 4 & 1.5 & 1.3 & AM0178\,--\,(NVAS)\\
  352.0       & 1/8 & 0.36 & 0.25 & 4 & 4.7 & 0.35 & AG247\,--\,(MJH)\\
  356.0       & 1/4 & 0.72 & 0.75 & 4 & 4.9 & 0.45x0.38 & AF0186\,--\,(NVAS)\\
  368.0       & 1/4 & 0.72 & 0.3 & 4 & 8.5 & 0.2 & AL0322\,--\,(NVAS)\\
  382.0       & 1/4 & 0.72 & 0.125 & 4 & 8.4 & 0.75 & PERL\,--\,(NED)\\
  388.0       & 1/8 & 0.94 & 0.25 & 4 & 4.9 & 0.47x0.36 & AC0149\,--\,(NVAS)\\
  401.0       & 1/8 & 0.51 & 0.1 & 4 & 8.4 & 0.27 & AP315\,--\,(MJH)\\
  (401b)      & 1/2 & 3.8 & 0.1 & 4 & 8.4 & 0.27 & AP315\,--\,(MJH)\\
  427.1       & 1/4 & 0.72 & 0.1 & 4 & 8.5 & 0.25 & AP331\,--\,(MJH)\\
  (427.1b)    & 1/2 & 3.8 & 0.1 & 4 & 8.5 & 0.25 & AP331\,--\,(MJH)\\
  432.0       & 1/8 & 0.51 & 2.0 & 2 & 4.9 & 0.40 & AB0454\,--\,(NVAS)\\
  433.0       & 1/4 & 0.72 & 0.1 & 2 & 8.5 & 0.25 & AB534\,--\,(MJH)\\
  437.0       & 1/4 & 1.0 & 4.0 & 4 & 4.9 & 1.2 & AV164\,--\,(TW)\\
  438.0       & 1/4 & 1.3 & 0.125 & 4 & 8.4 & 0.23 & AP315\,--\,(NED)\\
  441.0       & 1/8 & 0.51 & 0.25 & 4 & 4.9 & 0.35 & AF213\,--\,(MJH)\\
  442.0       & 2 & 5.8 & 0.5 & 2 & 1.4 & 7.5 & PEGG\,--\,(NED)\\
  449.0       & 2 & 8.1 & 0.125 & 4 & 1.5 & 4.0 & AK319\,--\,(CCC)\\
  (insert)    & 1/2 & 1.4 & 0.25 & 4 & 1.7 & 1.2 & PERL\,--\,(NVAS)\\
  455.0       & 1/8 & 0.51 & 0.5 & 4 & 4.9 & 0.40 & AP331\,--\,(MJH)\\
  469.1       & 1/4 & 0.72 & 2.0 & 4 & 4.9 & 1.7x1.1 & AR0123\,--\,(NVAS)\\
  470.0       & 1/4 & 0.72 & 2.0 & 4 & 8.4 & 1.5x1.3 & AL330\,--\,(TW)\\
\noalign{\smallskip}
\hline
\end{tabular}\\
Col. (1): The 3CR name.
Col. (2): The binning factor of the X-ray image (see \S~\ref{sec:xray} for more details).
Col. (3): the full width half maximum (FWHM) of the smoothing kernel chosen for the X-ray image.
Col. (4): The value of the lowest contour level of the radio map overlaid to the X-ray image.
Col. (5): The factor increase of the radio contours.
Col. (6): The radio frequency of the radio map used for the comparison with the X-ray image.
Col. (7): The half-power beam width (HPBW) of the reduced radio images. Single numbers are reported for circular beam.
Col. (8): The identification number of the observer program, as reported in the header of the raw u,v data downloaded from the VLA archive (see https://archive.nrao.edu/archive/nraodashelpj.html for more details).\\
Notes:\\
5. There may be a wcs problem with the coordinates of the radio image of order 1\arcsec; so the R.A./Dec. labels could be slightly off.
\end{table*} 

\newpage
\section{B: The Status of the \chn\ X-ray 3CR observations}
\label{sec:state}

Here we present the current status of the \chn\ X-ray observations for the entire 3CR catalog.
For each 3CR source we indicate the radio-to-optical classification indicating
FR\,I and FR\,II radio galaxies, according to the Fanaroff \& Riley criterion \citep{fanaroff74}; ii) quasars (i.e., QSRs);
Seyfert galaxies (Sy) and BL Lac objects (BL). 
We indicate as ``UND'' those sources which, lacking optical spectroscopy, remain unidentified.
Then the most updated value of the redshift $z$ is reported together with the luminosity distance $D_L$ and
we also used a ``cluster flag'' to label sources that belong to a known galaxy cluster.
Regarding the X-ray analysis, we report X-ray detections of radio components adopting the following symbols: 
k = jet knot; h = hotspot; l = lobe and xcl for sources that belong to a galaxy cluster detected in the X-rays.
Finally, the ``\chn\ flag'' indicates if the source was already observed by \chn.

\begin{table} 
\tiny
\caption{The current status of the 3CR \chn\ observations.}
\label{tab:main}
\begin{center}
\begin{tabular}{|rrrrrrr|}
\hline
3CR  & class & $z$ & D$_L$ & Cluster & X-ray     & Chandra \\  
name &          &       & Mpc   & flag    & detection & flag \\
\hline 
\noalign{\smallskip}
  2.0 & QSR & 1.037367 & 7252.26 & no & & yes\\
  6.1 & FRII & 0.8404 & 5577.63 & no & h & yes\\
  9.0 & QSR & 2.019922 & 16632.24 & no & k,l & yes\\
  11.1 & UND & ? &  & no &  & no\\
  13.0 & FRII & 1.351 & 10088.93 & no & h & yes\\
  14.0 & QSR & 1.469 & 11200.31 & no &  & yes\\
  14.1 & UND & ? &  & no &  & no\\
  15.0 & FRI & 0.073384 & 341.93 & no & k,l & yes\\
  16.0 & FRII & 0.405 & 2288.92 & no & h,l & yes\\
  17.0 & QSR & 0.219685 & 1126.33 & no & k & yes\\
  18.0 & FRII & 0.188 & 945.54 & no &  & yes\\
  19.0 & FRII & 0.482 & 2819.56 & yes & h & yes\\
  20.0 & FRII & 0.174 & 867.55 & no &  & yes\\
  21.1 & UND & ? &  & no &  & no\\
  22.0 & FRII & 0.936 & 6378.57 & no &  & yes\\
  27.0 & FRII & 0.184 & 923.17 & no &  & no\\
  28.0 & FRI & 0.195275 & 986.54 & yes & xcl & yes\\
  29.0 & FRI & 0.045031 & 205.48 & yes & k & yes\\
  31.0 & FRI & 0.017005 & 75.94 & yes & k & yes\\
  33.0 & FRII & 0.0597 & 275.49 & no & h & yes\\
  33.1 & FRII & 0.180992 & 906.44 & no &  & yes\\
  33.2 & UND & ? &  & no &  & no\\
  34.0 & FRII & 0.69 & 4368.53 & no &  & yes\\
  35.0 & FRII & 0.067013 & 310.8 & no &  & yes\\
  36.0 & QSR & 1.301 & 9624.8 & no &  & no\\
  40.0 & FRII & 0.018 & 80.46 & yes & xcl & yes\\
  41.0 & FRII & 0.795 & 5205.64 & no &  & yes\\
  42.0 & FRII & 0.395007 & 2221.96 & no &  & yes\\
  43.0 & QSR & 1.459 & 11105.41 & no &  & yes\\
  44.0 & FRII & 0.66 & 4135.83 & yes &  & yes\\
  46.0 & FRII & 0.4373 & 2508.42 & yes &  & yes\\
  47.0 & QSR & 0.425 & 2424.27 & no & h,l & yes\\
  48.0 & QSR & 0.367 & 2036.73 & no &  & yes\\
  49.0 & FRII & 0.621 & 3837.6 & no &  & yes\\
  52.0 & FRII & 0.29 & 1546.9 & yes & h & yes\\
  54.0 & FRII & 0.8274 & 5470.44 & no &  & yes\\
  55.0 & FRII & 0.7348 & 4721.63 & no &  & yes\\
  63.0 & FRII & 0.175 & 873.1 & no &  & yes\\
  61.1 & FRII & 0.18781 & 944.47 & no & h & yes\\
  65.0 & FRII & 1.176 & 8483.18 & no & h & yes\\
  66.0A & BL & ? &  & yes &  & yes\\
  66.0B & FRI & 0.021258 & 95.28 & yes & k & yes\\
  67.0 & FRII & 0.3102 & 1672.55 & no &  & yes\\
  68.1 & QSR & 1.238 & 9045.91 & no &  & yes\\
  68.2 & FRII & 1.575 & 12216.2 & no & h & yes\\
  69.0 & FRII & 0.458 & 2651.47 & no &  & no\\
  71.0 & Sy & 0.003793 & 16.72 & no &  & yes\\
  75.0 & FRI & 0.023153 & 103.9 & yes & xcl & yes\\
  76.1 & FRII & 0.032489 & 146.82 & no &  & yes\\
  78.0 & FRI & 0.028653 & 129.09 & no & k & yes\\
\noalign{\smallskip}
\hline
\end{tabular}\\
\end{center}
Col. (1): The 3CR name.
Col. (2): The radio-to-optical classification of the sources: FR\,I and FR\,II refer to the Fanaroff and Riley classification criterion for radio galaxies \citep{fanaroff74}; 
QSR stands for quasars; Sy for Seyfert galaxies and BL for BL Lac objects. We used the acronym UND for sources that are still unidentified; i.e.,  lacking of an optical spectroscopic observation.
Col. (3): Redshift $z$. We also verified in the literature (e.g., NED and/or SIMBAD databases) if new $z$ values were reported after the release of the 3CR catalog.
Col. (4): Luminosity Distance in Mpc. Cosmological parameters used to compute it are reported in \S~\ref{sec:intro}. 
Col. (5): The ``cluster flag'' indicates if the source is known to be associated with a catalogued cluster of galaxies or if there is significantly extended X-ray emission around the host galaxy; i.e., on scales of 100 kpc or greater.
Col. (6): In this column we report if the source has a radio component with an X-ray counterpart. We used the following labels: k = jet knot; h = hotspot; l = lobe.
We also indicated xcl if there is a galaxy cluster detected in the X-rays.
Col. (7): The ``\chn\ flag'' indicates if the source was already observed by \chn.
\end{table}

\begin{table} 
\tiny
\caption{The current status of the 3CR \chn\ observations.}
\begin{center}
\begin{tabular}{|rrrrrrr|}
\hline
3CR  & class & $z$ & D$_L$ & Cluster & X-ray     & Chandra \\  
name &  &     & Mpc   & flag    & detection & flag \\
\hline 
\noalign{\smallskip}
  79.0 & FRII & 0.255900 & 1339.62 & yes &  & yes\\
  83.1 & FRI & 0.025137 & 112.95 & yes & k,xcl & yes\\
  84.0 & FRI & 0.017559 & 78.45 & yes & xcl & yes\\
  86.0 & FRII & ? &  & no &  & no\\
  88.0 & FRI & 0.030221 & 136.32 & yes & k,xcl & yes\\
  89.0 & FRI & 0.1386 & 675.57 & yes & xcl & yes\\
  91.0 & UND & ? &  & no &  & no\\
  93.0 & QSR & 0.35712 & 1972.32 & no &  & yes\\
  93.1 & FRII & 0.243 & 1262.81 & yes &  & yes\\
  98.0 & FRII & 0.030454 & 137.4 & no &  & yes\\
  99.0 & Sy & 0.426 & 2431.07 & yes &  & yes\\
  103.0 & FRII & 0.33 & 1797.71 & no &  & yes\\
  105.0 & FRII & 0.089 & 419.33 & no & k,h & yes\\
  107.0 & FRII & 0.785 & 5124.51 & no &  & yes\\
  109.0 & FRII & 0.3056 & 1643.78 & no & h,l & yes\\
  111.0 & FRII & 0.0485 & 221.87 & no & k,h & yes\\
  114.0 & FRII & 0.815 & 5368.77 & no &  & yes\\
  119.0 & FRII & 1.023 & 7127.0 & no &  & no\\
  123.0 & FRII & 0.2177 & 1114.8 & yes & k & yes\\
  124.0 & FRII & 1.083 & 7653.06 & no &  & no\\
  125.0 & UND & ? &  & no &  & no\\
  129.0 & FRI & 0.0208 & 93.2 & yes & k,xcl & yes\\
  129.1 & FRI & 0.0222 & 99.56 & no &  & yes\\
  130.0 & FRI & 0.109 & 520.76 & no &  & yes\\
  131.0 & UND & ? &  & no &  & no\\
  132.0 & FRII & 0.214 & 1093.4 & yes &  & yes\\
  133.0 & FRII & 0.2775 & 1470.27 & no &  & yes\\
  134.0 & UND & ? &  & no &  & no\\
  135.0 & FRII & 0.12738 & 616.21 & yes &  & yes\\
  136.1 & FRII & 0.064 & 296.2 & no &  & yes\\
  137.0 & UND & ? &  & no &  & no\\
  138.0 & QSR & 0.759 & 4915.0 & no &  & yes\\
  139.2 & FRII & ? &  & no &  & no\\
  141.0 & UND & ? &  & no &  & no\\
  142.1 & FRII & 0.4061 & 2296.3 & no &  & yes\\
  147.0 & QSR & 0.545 & 3271.83 & no &  & yes\\
  152.0 & UND & ? &  & no &  & no\\
  153.0 & FRII & 0.2769 & 1466.59 & yes &  & yes\\
  154.0 & QSR & 0.58 & 3529.84 & no &  & yes\\
  158.0 & UND & ? &  & no &  & no\\
  165.0 & FRII & 0.2957 & 1582.19 & no &  & yes\\
  166.0 & FRII & 0.2449 & 1274.04 & no &  & yes\\
  169.1 & FRII & 0.633 & 3928.69 & no &  & yes\\
  171.0 & FRII & 0.2384 & 1235.67 & no &  & yes\\
  172.0 & FRII & 0.5191 & 3084.06 & no &  & yes\\
  173.0 & QSR & 1.035 & 7231.55 & no &  & no\\
  173.1 & FRII & 0.2921 & 1559.87 & yes & h,l & yes\\
  175.0 & QSR & 0.77 & 5003.31 & no &  & yes\\
  175.1 & FRII & 0.92 & 6242.98 & no &  & yes\\
  180.0 & FRII & 0.22 & 1128.16 & no &  & yes\\
\noalign{\smallskip}
\hline
\end{tabular}\\
\end{center}
\end{table}

\begin{table} 
\tiny
\caption{The current status of the 3CR \chn\ observations.}
\begin{center}
\begin{tabular}{|rrrrrrr|}
\hline
3CR  & class & $z$ & D$_L$ & Cluster & X-ray     & Chandra \\  
name &  &     & Mpc   & flag    & detection & flag \\
\hline 
\noalign{\smallskip}
  181.0 & QSR & 1.382 & 10378.89 & no & h & yes\\
  184.0 & FRII & 0.994 & 6875.53 & no & & yes\\
  184.1 & FRII & 0.1182 & 568.34 & yes &  & yes\\
  186.0 & QSR & 1.068634 & 7526.33 & yes & xcl & yes\\
  187.0 & FRII & 0.465 & 2700.19 & no & l & yes\\
  190.0 & QSR & 1.195649 & 8660.73 & no &  & yes\\
  191.0 & QSR & 1.956 & 15984.33 & no & k,l & yes\\
  192.0 & FRII & 0.059709 & 275.53 & yes &  & yes\\
  194.0 & FRII & 1.184 & 8555.38 & no &  & no\\
  196.0 & QSR & 0.871 & 5831.32 & no &  & yes\\
  196.1 & FRII & 0.198 & 1002.01 & no &  & yes\\
  197.1 & FRII & 0.128009 & 619.51 & yes &  & yes\\
  198.0 & FRII & 0.081474 & 381.9 & yes &  & yes\\
  200.0 & FRII & 0.458 & 2651.47 & yes & k,l & yes\\
  204.0 & QSR & 1.112 & 7909.99 & no &  & yes\\
  205.0 & QSR & 1.534 & 11821.39 & no &  & yes\\
  207.0 & QSR & 0.6808 & 4296.94 & no & k,l & yes\\
  208.0 & QSR & 1.111510 & 7905.63 & no &  & yes\\
  208.1 & QSR & 1.02 & 7100.98 & no &  & no\\
  210.0 & FRII & 1.169 & 8420.05 & no & h & yes\\
  212.0 & QSR & 1.048 & 7345.18 & no & h & yes\\
  213.1 & FRI & 0.19392 & 978.87 & yes & h & yes\\
  215.0 & QSR & 0.4121 & 2336.74 & no & k,l & yes\\
  217.0 & FRII & 0.8975 & 6053.2 & no &  & yes\\
  216.0 & QSR & 0.669915 & 4212.31 & no &  & yes\\
  219.0 & FRII & 0.174732 & 871.61 & yes & k,l & yes\\
  220.1 & FRII & 0.61 & 3754.31 & yes & xcl & yes\\
  220.2 & QSR & 1.157429 & 8316.07 & no &  & no\\
  220.3 & FRII & 0.68 & 4290.73 & no &  & yes\\
  222.0 & FRI & 1.339 & 9977.24 & no &  & no\\
  223.0 & FRII & 0.13673 & 665.6 & yes &  & yes\\
  223.1 & FRII & 0.107474 & 512.94 & no &  & yes\\
  225.0A & FRII & 1.565 & 12119.69 & no &  & yes\\
  225.0B & FRII & 0.58 & 3529.84 & no &  & yes\\
  226.0 & FRII & 0.8177 & 5390.92 & no &  & yes\\
  227.0 & FRII & 0.086272 & 405.71 & no & h & yes\\
  228.0 & FRII & 0.5524 & 3325.95 & no & h & yes\\
  230.0 & FRII & 1.487 & 11371.7 & no &  & no\\
  231.0 & FRI & 0.000677 & 2.97 & no &  & yes\\
  234.0 & FRII & 0.184925 & 928.33 & no & h & yes\\
  236.0 & FRII & 0.1005 & 477.44 & no &  & yes\\
  237.0 & FRII & 0.877 & 5881.45 & no &  & yes\\
  238.0 & FRII & 1.405 & 10594.89 & no &  & no\\
  239.0 & FRII & 1.781 & 14232.66 & no &  & no\\
  241.0 & FRII & 1.617 & 12622.99 & no &  & yes\\
  244.1 & FRII & 0.428 & 2444.69 & yes &  & yes\\
  245.0 & QSR & 1.027872 & 7169.36 & no & k & yes\\
  247.0 & FRII & 0.7489 & 4834.0 & yes &  & yes\\
  249.0 & QSR & 1.554 & 12013.69 & no &  & no\\
  249.1 & QSR & 0.3115 & 1680.71 & no &  & yes\\
  250.0 & FRII & ? &  & no &  & no\\
\noalign{\smallskip}
\hline
\end{tabular}\\
\end{center}
\end{table}

\begin{table} 
\tiny
\caption{The current status of the 3CR \chn\ observations.}
\begin{center}
\begin{tabular}{|rrrrrrr|}
\hline
3CR  & class & $z$ & D$_L$ & Cluster & X-ray     & Chandra \\  
name &  &     & Mpc   & flag    & detection & flag \\
\hline 
\noalign{\smallskip}
  252.0 & FRII & 1.1 & 7803.57 & no &  & yes\\
  254.0 & QSR & 0.736619 & 4736.12 & no & h & yes\\
  255.0 & QSR & 1.355 & 10126.27 & no &  & no\\
  256.0 & FRII & 1.819 & 14610.18 & no &  & yes\\
  257.0 & QSR & 2.474 & 21340.67 & no &  & no\\
  258.0 & FRI & 0.165 & 818.03 & yes &  & yes\\
  263.0 & QSR & 0.646 & 4028.06 & no & h & yes\\
  263.1 & FRII & 0.824 & 5442.55 & no &  & yes\\
  264.0 & FRI & 0.021718 & 97.37 & yes & k & yes\\
  265.0 & FRII & 0.811 & 5336.02 & no & h,l & yes\\
  266.0 & FRII & 1.275 & 9384.99 & no &  & yes\\
  267.0 & FRII & 1.14 & 8159.88 & no &  & yes\\
  268.1 & FRII & 0.97 & 6668.89 & no & h & yes\\
  268.2 & FRII & 0.362 & 2004.12 & yes & h & yes\\
  268.3 & FRII & 0.37171 & 2067.61 & no & & yes\\
  268.4 & QSR & 1.402200 & 10568.59 & no &  & yes\\
  270.0 & FRI & 0.007378 & 32.63 & yes & k & yes\\
  270.1 & QSR & 1.528432 & 11767.94 & no &  & yes\\
  272.0 & FRII & 0.944 & 6446.64 & no &  & yes\\
  272.1 & FRI & 0.003392 & 14.94 & yes & k & yes\\
  273.0 & QSR & 0.158339 & 781.73 & no & k & yes\\
  274.0 & FRI & 0.004283 & 18.89 & yes & k,xcl & yes\\
  274.1 & FRII & 0.422 & 2403.91 & no &  & yes\\
  275.0 & FRII & 0.48 & 2805.51 & yes &  & yes\\
  275.1 & QSR & 0.5551 & 3345.78 & no & k,h,l & yes\\
  277.0 & FRI & 0.414 & 2349.59 & no &  & yes\\
  277.1 & QSR & 0.31978 & 1732.99 & no & & yes\\
  277.2 & FRII & 0.766 & 4971.13 & no &  & yes\\
  277.3 & FRII & 0.085336 & 401.05 & no &  & yes\\
  280.0 & FRII & 0.996 & 6892.84 & yes & k,h,l & yes\\
  280.1 & QSR & 1.667065 & 13110.8 & no & l & no\\
  284.0 & FRII & 0.239754 & 1243.68 & yes &  & yes\\
  285.0 & FRII & 0.0794 & 371.64 & no &  & yes\\
  286.0 & QSR & 0.849934 & 5656.31 & no &  & yes\\
  287.0 & QSR & 1.055 & 7406.56 & no &  & yes\\
  287.1 & FRII & 0.215567 & 1102.45 & no & h & yes\\
  288.0 & FRI & 0.246 & 1280.56 & yes & xcl & yes\\
  288.1 & QSR & 0.96296 & 6608.61 & no &  & yes\\
  289.0 & FRII & 0.9674 & 6646.6 & no &  & yes\\
  292.0 & FRII & 0.71 & 4525.37 & no &  & yes\\
  293.0 & FRI & 0.045034 & 205.5 & no &  & yes\\
  293.1 & FRII & 0.709 & 4517.5 & no &  & yes\\
  294.0 & FRII & 1.779 & 14212.84 & yes & h,xcl & yes\\
  295.0 & FRII & 0.4641 & 2693.92 & yes & h,xcl & yes\\
  296.0 & FRI & 0.024704 & 110.97 & no & k & yes\\
  297.0 & QSR & 1.4061 & 10605.23 & no &  & no\\
  298.0 & QSR & 1.438120 & 10907.49 & no & & yes\\
  299.0 & FRII & 0.367 & 2036.73 & yes & h & yes\\
  300.0 & FRII & 0.27 & 1424.56 & no &  & yes\\
\noalign{\smallskip}
\hline
\end{tabular}\\
\end{center}
\end{table}

\begin{table} 
\tiny
\caption{The current status of the 3CR \chn\ observations.}
\begin{center}
\begin{tabular}{|rrrrrrr|}
\hline
3CR  & class & $z$ & D$_L$ & Cluster & X-ray     & Chandra \\  
name &  &     & Mpc   & flag    & detection & flag \\
\hline 
\noalign{\smallskip}
  300.1 & FRII & 1.15885 & 8328.86 & no &  & no\\
  303.0 & FRI & 0.141186 & 689.35 & yes & k & yes\\
  303.1 & FRI & 0.2704 & 1426.99 & no &  & yes\\
  305.0 & FRII & 0.041639 & 189.56 & no &  & yes\\
  305.1 & FRII & 1.132 & 8088.23 & no &  & no\\
  306.1 & FRII & 0.441 & 2533.89 & yes &  & yes\\
  309.1 & QSR & 0.905 & 6116.25 & no &  & yes\\
  310.0 & FRI & 0.0538 & 247.11 & yes & xcl & yes\\
  314.1 & FRI & 0.1197 & 576.16 & yes &  & yes\\
  313.0 & FRII & 0.461000 & 2672.37 & yes & h,xcl & yes\\
  315.0 & FRI & 0.1083 & 517.17 & yes &  & yes\\
  317.0 & FRI & 0.034457 & 155.96 & yes & xcl & yes\\
  318.0 & FRII & 1.574 & 12206.53 & yes &  & yes\\
  318.1 & FRI & 0.045311 & 206.8 & yes & xcl & yes\\
  319.0 & FRII & 0.192 & 968.03 & yes &  & yes\\
  320.0 & FRII & 0.342 & 1874.59 & yes & xcl & yes\\
  321.0 & FRII & 0.0961 & 455.08 & no & h & yes\\
  322.0 & FRII & 1.681 & 13247.25 & no &  & no\\
  323.0 & FRII & 0.679 & 4282.93 & no &  & yes\\
  323.1 & QSR & 0.2643 & 1390.12 & yes &  & yes\\
  324.0 & FRII & 1.2063 & 8757.25 & yes & h & yes\\
  325.0 & FRII & 1.135 & 8115.06 & no & h & yes\\
  326.0 & FRII & 0.0895 & 421.83 & no &  & yes\\
  326.1 & FRII & 1.825 & 14669.88 & no &  & no\\
  327.0 & FRII & 0.1048 & 499.28 & yes & h & yes\\
  327.1 & FRI & 0.462 & 2679.32 & no & k & yes\\
  330.0 & FRII & 0.55 & 3308.36 & yes & h & yes\\
  332.0 & FRII & 0.151019 & 742.01 & yes &  & yes\\
  334.0 & QSR & 0.5551 & 3345.78 & no & k,l & yes\\
  336.0 & QSR & 0.926542 & 6298.25 & no &  & yes\\
  337.0 & FRII & 0.635 & 3943.96 & yes &  & yes\\
  338.0 & FRI & 0.030354 & 136.94 & yes & xcl & yes\\
  340.0 & FRII & 0.7754 & 5046.9 & no &  & yes\\
  341.0 & FRII & 0.448 & 2582.06 & no & k & yes\\
  343.0 & QSR & 0.988 & 6823.74 & no &  & yes\\
  343.1 & FRII & 0.75 & 4842.79 & no &  & yes\\
  345.0 & QSR & 0.5928 & 3625.15 & no & k & yes\\
  346.0 & FRI & 0.162012 & 801.73 & yes & k & yes\\
  348.0 & FRI & 0.155 & 763.55 & yes & xcl & yes\\
  349.0 & FRII & 0.205 & 1041.79 & no & h & yes\\
  351.0 & FRII & 0.37194 & 2069.13 & no & h & yes\\
  352.0 & FRII & 0.8067 & 5300.91 & no &  & yes\\
  353.0 & FRII & 0.030421 & 137.25 & no & k & yes\\
  356.0 & FRII & 1.079 & 7617.8 & no &  & yes\\
  357.0 & FRII & 0.166148 & 824.31 & yes &  & yes\\
  368.0 & FRII & 1.131 & 8079.29 & no & & yes\\
  371.0 & BL & 0.051 & 233.74 & no & k & yes\\
  379.1 & FRII & 0.256 & 1340.22 & no &  & yes\\
  380.0 & QSR & 0.692 & 4384.16 & no & k & yes\\
  381.0 & FRII & 0.1605 & 793.51 & no &  & yes\\
  382.0 & FRII & 0.05787 & 266.65 & no &  & yes\\
\noalign{\smallskip}
\hline
\end{tabular}\\
\end{center}
\end{table}

\begin{table} 
\tiny
\caption{The current status of the 3CR \chn\ observations.}
\begin{center}
\begin{tabular}{|rrrrrrr|}
\hline
3CR  & class & $z$ & D$_L$ & Cluster & X-ray     & Chandra \\  
name &  &     & Mpc   & flag    & detection & flag \\
\hline 
\noalign{\smallskip}
  386.0 & FRI & 0.016885 & 75.39 & no &  & yes\\
  388.0 & FRII & 0.0917 & 432.88 & yes & xcl & yes\\
  389.0 & UND & ? &  & no &  & no\\
  390.0 & UND & ? &  & no &  & no\\
  390.3 & FRII & 0.0561 & 258.14 & no & k,h & yes\\
  394.0 & UND & ? &  & no &  & no\\
  399.1 & FRII & ? &  & no &  & no\\
  401.0 & FRII & 0.2011 & 1019.64 & yes & xcl & yes\\
  402.0 & FRI & 0.025948 & 116.67 & yes & k & yes\\
  403.0 & FRII & 0.059 & 272.1 & no & k,h & yes\\
  403.1 & FRII & 0.0554 & 254.77 & yes &  & yes\\
  405.0 & FRII & 0.056075 & 258.01 & yes & h,xcl & yes\\
  409.0 & FRII & ? &  & no &  & no\\
  410.0 & FRII & 0.2485 & 1295.4 & no &  & yes\\
  411.0 & FRII & 0.467 & 2714.14 & no &  & yes\\
  415.2 & UND & ? &  & no &  & no\\
  418.0 & QSR & 1.686 & 13296.15 & no &  & no\\
  424.0 & FRI & 0.126988 & 614.15 & yes &  & yes\\
  427.1 & FRII & 0.572 & 3470.35 & yes & l,xcl & yes\\
  428.0 & UND & ? &  & no &  & no\\
  430.0 & FRII & 0.055545 & 255.47 & yes &  & yes\\
  431.0 & UND & ? &  & no &  & no\\
  432.0 & QSR & 1.785 & 14272.26 & no &  & yes\\
  434.0 & FRII & 0.322 & 1746.99 & yes &  & yes\\
  433.0 & FRI & 0.1016 & 483.01 & no &  & yes\\
  435.0 & FRII & 0.471 & 2742.13 & no &  & yes\\
  436.0 & FRII & 0.2145 & 1096.28 & no & h & yes\\
  437.0 & FRII & 1.48 & 11305.11 & no & h & yes\\
  438.0 & FRII & 0.29 & 1546.9 & yes & xcl & yes\\
  441.0 & FRII & 0.708 & 4509.63 & no &  & yes\\
  442.0 & FRI & 0.0263 & 118.28 & yes & xcl & yes\\
  445.0 & FRII & 0.055879 & 257.07 & yes & h & yes\\
  449.0 & FRI & 0.017085 & 76.3 & yes & xcl & yes\\
  452.0 & FRII & 0.081100 & 380.05 & no & h,l & yes\\
  454.0 & QSR & 1.757 & 13995.04 & no &  & no\\
  454.1 & FRII & 1.841 & 14829.49 & yes &  & no\\
  454.2 & UND & ? &  & no &  & no\\
  454.3 & QSR & 0.859 & 5731.6 & no & k & yes\\
  455.0 & QSR & 0.543 & 3257.27 & no &  & yes\\
  456.0 & FRII & 0.233 & 1203.84 & no &  & yes\\
  458.0 & FRII & 0.289 & 1540.74 & yes & h & yes\\
  459.0 & FRII & 0.22012 & 1128.85 & no & l & yes\\
  460.0 & FRII & 0.268 & 1412.45 & yes &  & yes\\
  465.0 & FRI & 0.030221 & 136.32 & yes & k,xcl & yes\\
  468.1 & UND & ? &  & no &  & no\\
  469.1 & FRII & 1.336 & 9949.27 & no &  & yes\\
  470.0 & FRII & 1.653 & 12973.42 & no & h & yes\\
\noalign{\smallskip}
\hline
\end{tabular}\\
\end{center}
\end{table}

\begin{figure}
\includegraphics[keepaspectratio=true,scale=0.4]{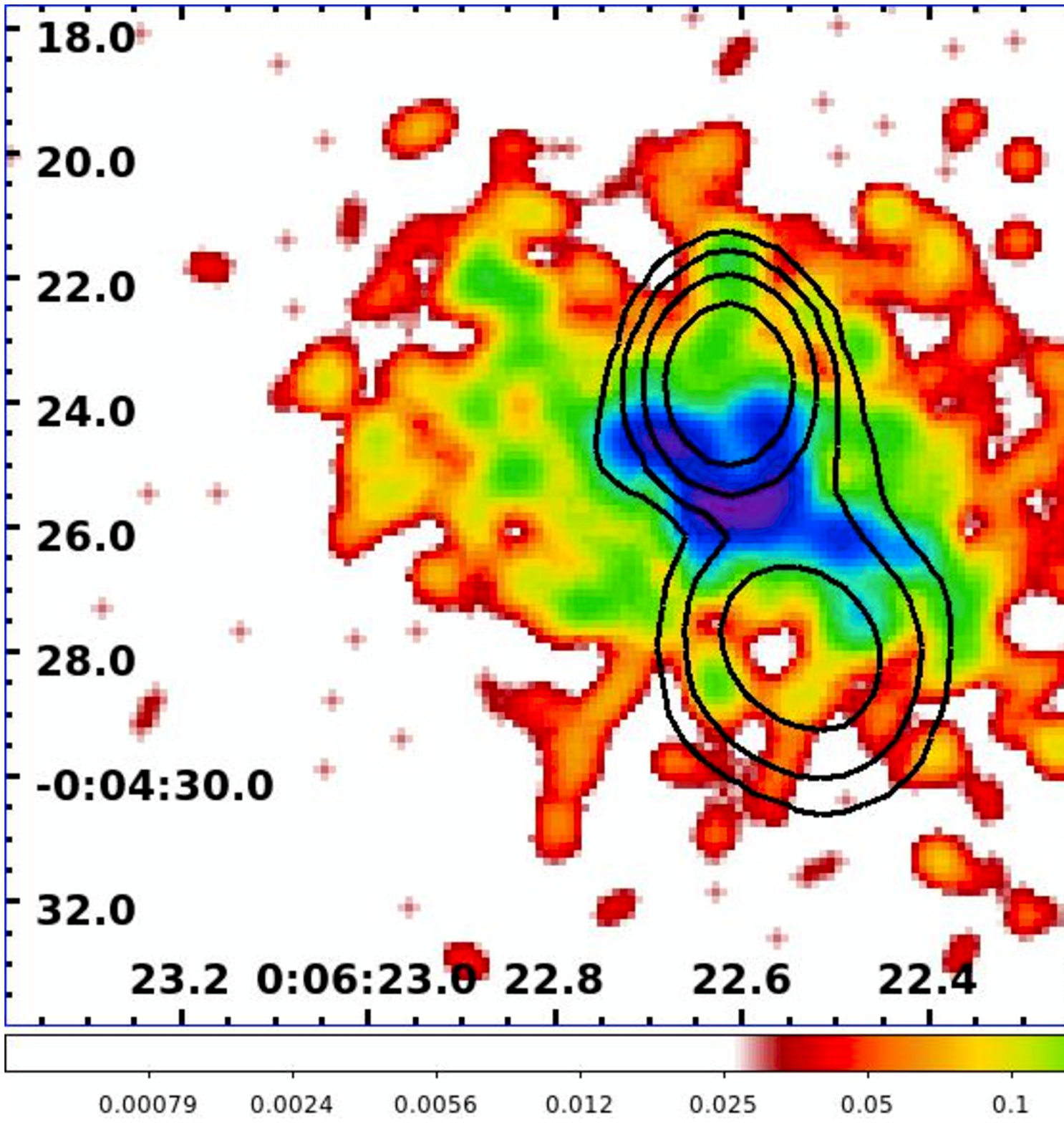}
\caption{The X-ray image corresponding to the Chandra observation (Tables \ref{tab:log1} and \ref{tab:log2}) with contours of radio brightness superposed.  The image is re-binned to change the pixel size and is smoothed with a Gaussian function.  The underlying color bar shows the X-ray brightness in units of counts per pixel. Radio contours are logarithmically spaced.  All relevant parameters for each source are given  in Tables \ref{tab:captions1} and \ref{tab:captions2}.}
\label{fig:3c2app}
\end{figure}

\end{document}